\documentclass[twocolumn]{aa}

\date{Received 13 August 2025 / Accepted 20 November 2025}

\usepackage[varg]{txfonts}
\usepackage[utf8]{inputenc}
\usepackage[english]{babel}
\usepackage{tabularx}

\usepackage{setspace}

\usepackage{indentfirst}

\usepackage{enumitem}
\setlist[itemize]{noitemsep, topsep=0pt}

\usepackage{graphicx}
\graphicspath{{./}}

\usepackage{wrapfig}
\usepackage{subfigure}
\usepackage[dvipsnames]{xcolor}

\usepackage{hyperref}
\hypersetup{
    colorlinks,
    urlcolor=blue,
    linkcolor=blue,
    citecolor=blue
}

\usepackage{csquotes}

\usepackage{listings}
\usepackage{courier}\definecolor{bluekeywords}{rgb}{0,0,1}
\definecolor{greencomments}{rgb}{0,0.5,0}
\definecolor{redstrings}{rgb}{0.64,0.08,0.08}
\definecolor{types}{rgb}{0.17,0.57,0.68}
\lstdefinestyle{py}
{
    language=Python,
    frame=l,
    framesep=5pt,
    captionpos=b,
    numbers=left,
    numberstyle=\tiny,
    showspaces=false,
    showtabs=false,
    breaklines=true,
    showstringspaces=false,
    breakatwhitespace=true,
    commentstyle=\color{greencomments},
    keywordstyle=\color{bluekeywords},
    stringstyle=\color{redstrings},
    basicstyle=\footnotesize\ttfamily,
}
\newcommand\Tstrut{\rule{0pt}{2.6ex}}       \newcommand\Bstrut{\rule[-0.9ex]{0pt}{0pt}} \newcommand{\TBstrut}{\Tstrut\Bstrut} \defcitealias{2025A&A...700A..53M}{Paper I}
\usepackage{natbib}
\bibpunct{(}{)}{;}{a}{}{,} 
\begin{document}

\title{Young M dwarfs flare activity model: towards better exoplanetary atmospheric characterisation}
\author{E. Mamonova\inst{1}, A. F. Kowalski\inst{2,3,4}, K. Herbst\inst{1}, S. Wedemeyer\inst{5,6}, S. C. Werner\inst{1}}

\institute{Centre for Planetary Habitability (PHAB), University of Oslo, 0315 Oslo, Norway
\and National Solar Observatory, University of Colorado Boulder, 3665 Discovery Drive, Boulder, CO 80303, USA
\and Department of Astrophysical and Planetary Sciences, University of Colorado, Boulder, 2000 Colorado Ave, CO 80305, USA
\and Laboratory for Atmospheric and Space Physics, University of Colorado Boulder, 3665 Discovery Drive, Boulder, CO 80303, USA
\and Rosseland Centre for Solar Physics, University of Oslo, 0315 Oslo, Norway
\and Institute of Theoretical Astrophysics, University of Oslo, 0315 Oslo, Norway}

\abstract{\textit{Context.} Stellar flares can significantly influence the atmospheres and habitability of orbiting exoplanets, especially around young and active M dwarfs. Understanding the temporally and spectrally resolved activity of such stars is essential for assessing their impact on planetary environments. \\
\textit{Aims.} We aim to examine in detail state-of-art concepts of flare models to identify what is missing in our understanding of energy deposition during the flare event. By comparing synthetic and observed flare spectra, we seek to determine the modelling frameworks best suited to represent flare energetics and spectral far-ultraviolet features while providing a foundation for investigating flare impacts on exoplanet atmospheres.\\
\textit{Methods.} In this work, we built the Young M Dwarfs Flare (YMDF) model utilising the combination of radiative-hydrodynamic (RHD) stellar atmosphere models with a high and low-energy electron beam and corresponding synthetic observables. These models are based on physical principles and were validated with solar and stellar observations.\\
\textit{Results.} The newly developed YMDF model reproduces the observed continuum rise in both the TESS photometric band and the FUV-A spectral range. Furthermore, the flare distributions generated within this framework show consistency with those observed in our sample of stars.\\
\textit{Conclusions.} We have developed the YMDF model as a tool to reproduce the time-dependent spectra of flaring young M~dwarfs, providing a physically motivated description of their spectral and temporal evolution during flare events.

}
\titlerunning{The Young M Dwarf Flare (YMDF) model.}
\authorrunning{Mamonova et al.}
\maketitle

\section{Introduction}
\label{sec:introduction}

Exoplanet science is rapidly advancing towards the discovery of Earth-like planets (i.e. in terms of size and mass), situated within the liquid-water habitable zone \citep{Owen1980}. Since the early 2010s, studies by \citet{2013ApJ...762...41G} and \citet{2015ApJ...807...45D} have shown that planets in the liquid-water habitable zone are more common around M dwarfs. Consequently, M dwarfs have emerged as the most prevalent type of star known to host Earth-like planets. It is estimated that approximately 40\% of M dwarfs could harbour one or more planets within their habitable zone \citep{2013A&A...549A.109B,2015ApJ...814..130M,2015ApJ...798..112M}. The relative abundance of M stars makes them prime targets in the search for habitable worlds.

The question of whether M stars can sustain habitable planets, however, remains open, partly due to their strong chromospheric activity and frequent flare production \citep{2010AsBio..10..751S}. While flaring activity may alter and erode planetary atmospheres \citep{2007AsBio...7..185L,2010AsBio..10..751S,2019AsBio..19...64T}, it may also provide the necessary ultra-violet (UV) radiation for prebiotic chemistry processes \citep{2017ApJ...843..110R,2021AsBio..21.1099R}.

The presence of self-sustained magnetic fields in low-mass stars significantly impacts their upper atmospheric structure and high-energy radiative environments. Various heating processes including shock dissipation of acoustic waves generated in stellar surface convection zones \citep{1996SSRv...75..453N} and magnetic reconnection \citep{2006SoPh..234...41K} cause these low-mass stars to exhibit significant temperature inversions in their outer atmospheres. The increased power and frequency of M dwarf flares may be attributed to their turbulent magnetic dynamos and strong magnetic fields. It remains uncertain whether the mechanisms that produce M dwarf flares are truly analogous to those of the Sun. It is believed that stars below approximately 0.35 M$_\odot$ become fully convective, lacking a radiative zone and consequently a tachocline \citep{1997A&A...327.1039C}. This transition to full convection may involve a shift from a Solar-type magnetic dynamo governed by rotation to a turbulent dynamo potentially less dependent on rotation \citep{2022A&A...662A..41R}. Despite extensive research on stellar chromospheres and coronae, the nature of these layers, the underlying processes that generate them, and their evolution remain poorly understood, particularly for M-dwarf stars \citep{2021ApJ...911..111P}.

Flares, manifesting as broadband emission ranging from the near-UV (NUV) through optical wavelengths and, in some cases, extending into the far-UV (FUV) and near-infrared (NIR), have been extensively studied in recent years. This progress is largely due to space-based dedicated photometric monitoring missions such as Kepler \citep{2010Sci...327..977B}, its extension K2 \citep{2014PASP..126..398H}, and the ongoing Transiting Exoplanet Survey Satellite (TESS, \citealt{2015JATIS...1a4003R}).

Flares and associated energetic particle events have the potential to alter the chemistry and climate of exoplanetary atmospheres, particularly those with Earth-like characteristics \citep{2007AsBio...7..185L,2010AsBio..10..751S,2016ApJ...830...77V,2019AsBio..19...64T,2019A&A...631A.101H,2024ApJ...961..164H}. However, these analyses were primarily based on observations of a limited number of well-characterized M dwarf flares at FUV wavelengths and extrapolations from solar and M dwarf flare activity observations at optical wavelengths, since there are significant challenges associated with direct UV and X-ray observations. Below the hydrogen ionisation edge at 912 $\AA$, stellar emission is absorbed by the interstellar medium, making UV and X-ray light observable only from space.

The question of whether M dwarfs can host habitable planets has attracted significant attention recently. For a detailed review of the numerous factors influencing the habitability of planets orbiting G, K, and M dwarfs, see \citet{2019BAAS...51c.564A,2024ApJ...960...62E}.

M dwarf flares are, on average, more energetic than solar flares \citep{1991ApJ...378..725H,2016ApJ...829...23D,2017ApJ...851...91N}. Typical white-light energies of dMe flares are $10^{30}$-$10^{32}$ erg \citep{2014ApJ...797..121H} with occasional events exceeding $10^{34}$ erg in the U-band \citep{2015SoPh..290.3487K}. The areal extent of flare footpoints cannot be measured directly, and hard X-ray emission is far too faint except during the largest dMe flares \citep{2010ApJ...721..785O}. Although \citet{2015SoPh..290.3487K} did not determine if nonthermal (NT) electron fluxes on M dwarfs are indeed as high as $10^{13}$ erg cm$^{-2}$ s$^{-1}$, typical values of $\sim 10^{11}$  erg cm$^{-2}$ s$^{-1}$ in strong solar flares cannot explain the properties of white-light emission during dMe flares. Thus, it is plausible to assume that the energy flux is higher.

Radiative-hydrodynamic models (RHD, \citealt{2005ApJ...630..573A,2017ApJ...837..125K}) grounded in physical principles have become essential for understanding the dynamic processes in stellar atmospheres. By coupling radiation and hydrodynamics, these models allow for self-consistent modelling that captures the complex interplay between energy transport, atmospheric structure, and observable phenomena such as flares. Validated against both solar and stellar observations, RHD models provide a robust framework interpreting a wide range of stellar activity and its signatures across different spectral regimes when subsequently used in radiative transfer calculations to produce observable signatures \citep{2019ApJ...871L..26F,2022FrASS...934458K}.

Spectroscopic studies \citep{2019ApJ...871L..26F,2023ApJ...959...64H} alongside broadband analyses \citep{2024ApJ...971...24P,2023ApJ...944....5B} have begun incorporating multicomponent models spanning from FUV through to NIR. The multi-wavelength approach of fitting the combinations of the RHD models, combining the TESS optical data and the FUV observations from the Cosmic Origins Spectrograph on the Hubble Space Telescope (HST-COS), provides a comprehensive view of magnetic activity in young M dwarfs.

In this paper, we discuss a newly developed Young M Dwarf Flare model (YMDF) to study exoplanetary atmospheres in the vicinity of young and active M dwarfs. The paper is organised as follows: in Section \ref{sec:methods}  we describe our methods for developing the model components, and the flare sample in TESS and FUV used for evaluating synthetic light curves and spectra. In Section \ref{sec:results} we present the results obtained using the YMDF model and analyse the synthetic stellar activity datasets. A comparison of our proposed model and previous modelling efforts is discussed in Section \ref{sec:discussion}.

\section{Methodology}
\label{sec:methods}

\subsection{Stellar sample}

In \citet{2025A&A...700A..53M} (hereafter, \citetalias{2025A&A...700A..53M}), we assembled a sample of 493 young and field M-K dwarfs and analysed their activity patterns, validating 86\,714 flare events in TESS and Kepler data and 52 events in HST-COS FUV data. In this work, we utilised a portion of this sample, specifically the observational data of stars exhibiting flaring activity in both FUV and TESS bandpasses (see Table \ref{tab:1}).

\begin{table}
\scriptsize
\caption{\label{tab:1}The sample of young active M-K dwarfs}
\centering
\resizebox{0.47\textwidth}{!}{\begin{tabularx}{11.5cm}{l c c c c c c c c } \hline
ID& SpT& d (pc) & P$_\mathrm{rot}$& T$_\mathrm{eff}$& R$_\star$/R$_\odot$& Age (Myr)& N$_\mathrm{FUV}$ & N$_\mathrm{TESS}$\TBstrut\\
\hline

2MASS J11173700-7704381 &M0.5 &188.875 & –$^{*}$ & 3778 &1.18 &5.0& 2&54  \TBstrut\\
AU Mic &M1 &9.714&4.8& 3835& 0.75& 24.0&13& 20 \TBstrut\\
2MASS J18141047-3247344$^{**}$&K5.0& 71.483&1.71&3512& 1.05&24.0&2&14\TBstrut\\
2MASS J02365171-5203036& M2 & 38.839&0.74&3626&0.74 &42.0 &4&73\TBstrut\\
2MASS J01521830-5950168&M2&39.594 &6.5&3626& 0.78&45.0& 2&5\TBstrut\\
HIP 107345 &M1 & 46.41&4.5 & 3837& 0.82& 45.0& 2&31 \TBstrut\\
HIP 1993 &M0 &44.148&4.3 &4053&0.71 &45.0 & 1 &30\TBstrut\\
2MASS J03315564-4359135& M0 & 45.11&2.9& 4491& 0.77&45.0&1&8  \TBstrut\\
2MASS J02001277-0840516& M2.8 &36.787&2.28& 3345&0.66&45.0 &1& 5 \TBstrut\\
2MASS J22025453-6440441&M2& 43.67&0.43&3044& 0.57&45.0& 1&48\TBstrut\\
Karmn J07446+035& M4.5& 5.9888&2.78&3099& 0.420&50.0 &3&41\TBstrut\\

\end{tabularx}}

   \tablefoot{Stellar parameters are from \citetalias{2025A&A...700A..53M} and available in detail via the CDS\footnotemark.
   N$_\mathrm{FUV}$ and N$_\mathrm{TESS}$ are numbers of flare events participating in the model fitting for observations in FUV and TESS, respectively. $^*$The rotation period was flagged as uncertain. For comparison, the average rotation period of stars in the Chamaeleon I moving group, which has an age of approximately 5 Myr \citep{2007ApJS..173..104L,2015A&A...575A...4F}, and which contributes to the sample described in \citetalias{2025A&A...700A..53M}, is between 4 and 6 days. $^{**}$Spectroscopic binary of two K dwarfs}.
\end{table}
\footnotetext{https://cdsarc.cds.unistra.fr/viz-bin/cat/J/A+A/700/A53}
\subsection{Stellar atmosphere models during a flare event}
\label{subsec:steatmodel}

Stellar flares are believed to originate from processes similar to those observed on the Sun. In solar flares, non-thermal electrons generate hard X-ray sources that are co-temporal and co-spatial with white-light radiation at the foot-points of reconnected magnetic loops. Thus, electron beam models, analogous to the ones used in this study, provide a physically grounded representation of the energy input in flares. The beam approach with RHD simulations models key stellar atmospheric properties such as opacities, flows, heating rates, and non-equilibrium radiation. RADYN assumes an impulsive flare phase driven by a downward-propagating charged particle beam accelerated in the corona, rapidly heating and expanding the lower atmosphere (i.e., chromospheric evaporation, \citealt{2015ApJ...809..104A}). This allows flare evolution modelling from impulsive to decay phases. \citet{2024ApJ...969..121K,2017ApJ...837..125K} successfully reproduced observed flare properties including optical and NUV continuum and Balmer emission lines, validating the electron beam framework as a robust stellar flare model foundation.

To model stellar flare heating, one-dimensional time-dependent model employ a power-law distribution of accelerated electrons \citep{2024ApJ...969..121K}, coupled with the RADYN radiative-hydrodynamic code \citep{1994chdy.conf...47C,1995ApJ...440L..29C,1997ApJ...481..500C}. RADYN self-consistently solves the equations of mass, momentum, and internal energy conservation, as well as non-equilibrium level populations, charge conservation, and radiative transfer \citep{1994chdy.conf...47C}. The non-local thermodynamic equilibrium (non-LTE) radiative transfer problem is addressed within RADYN using the accelerated $\Lambda$-iteration techniques developed by \citet{1981ApJ...249..720S} and \citet{1985JCoPh..59...56S}.
The electron beam enables significant heating at deeper atmospheric levels. By adjusting the parameters of the electron distribution, such models achieve optical blackbody temperatures of approximately 10,000 K in the emergent radiative flux spectra, consistent with observational data. Parameters that describe the electron beam power-law distribution, such as the low-energy cutoff (E$_c$), power-law index ($\delta$), and energy flux, govern the energy deposition and heating depth during flares.

\citet{2024ApJ...969..121K} suggested time-dependent stellar flare models of deep atmospheric heating in M dwarfs. In their grid the constant group includes models with large E$_c$ and constant flux injections reaching 10$^{13}$ erg cm$^{-2}$ s$^{-1}$, producing intense heating and continuum emission. The auxiliary group expands the grid with models that better reproduce observed features than the main group, comprised of impulsive beam models. The auxiliary models can have a lower E$_c$ and a hard power-law index $\delta$=2.5, representing a relatively low flux but energetic beam, reaching peak flux near 10$^{12}$ erg cm$^{-2}$ s$^{-1}$ in a pulsed ramp up/down injection, analogous to solar flare hard X-ray bursts characterised by rapid increases followed by gradual decreases in intensity.

\citet{2025ApJ...978...81K} found that only the most energetic electron beam heating model can reproduce the rising slope of flares they observed within the NUVA wavelength range, and in their framework, a lower E$_c$ provided an adequate explanation for the overall properties of the full UV continuum. They studied a recent super-flare in CR Dra, an M dwarf with properties broadly similar to AU Mic, although CR Dra has an uncertain age and a rotation period twice as fast. Thus, we adopted their approach: a linear combination of the two atmospheric models during a flare event, each scaled by its respective coefficient. The first atmospheric model (from the auxiliary group), m2F12-37-2.5, is characterised by a low flux density, an electron beam of 2$\times$10$^{12}$ erg s$^{-1}$ cm$^{-2}$, a smaller low-energy cut-off E$_c$=37 keV, and a hard power-law index $\delta$=2.5. The second, high flux density model (from the constant group), cF13-500-3, has an energy flux density of of 10$^{13}$ erg s$^{-1}$ cm$^{-2}$, a low-energy cut-off of E$_c$=500 keV, and a number-flux power-law index of $\delta$=3 above this low-energy cut-off. The physical origin of these two models can be attributed to fainter ribbons (associated with the low-flux m2F12-37-2.5 model) and bright kernel/kernels (associated with the high-flux cF13-500-3 model) in the flaring portion of the stellar surface, analogous to those observed in solar flare studies (e.g., \citealt[Fig. 6]{2022FrASS...934458K}).

To obtain the flux density of the models during flare events, we use the RH 1.5D code \citep{2015A&A...574A...3P}, which solves the non-LTE radiative transfer problem. The code employs the multi-level accelerated $\Lambda$-iteration method developed by \citet{2001ApJ...557..389U}, building on the approach of \citet{1991A&A...245..171R} to handle partial frequency redistribution (PRD) effects.
To efficiently solve the complex radiative transfer and atomic rate equations, RH 1.5D applies advanced numerical techniques that speed up convergence and improve solution stability, along with simplifying the system of the rate equations that govern atomic level populations, enabling the treatment of spectral line blends.

In addition to supporting overlapping radiative transitions, the RH 1.5D code solves the rate equations for multiple atoms simultaneously, consistently treating any overlapping transitions \citep{2015A&A...574A...3P}. It provides flexibility in the treatment of atoms by allowing transitions and continua to be computed either in non-LTE for active atoms (i.e.,  populations of atoms treated in detail and updated according to non-LTE radiative transfer) or under the assumption of LTE for passive atoms, treating them as opacity sources.

To run RH 1.5D, spectral setups have to be determined. We used 6-level hydrogen atom as the only non-LTE atom, and use the LTE conditions for Ca~II, Mg~II, Si, Al, Fe, He, N, Na, S atoms (hereafter, the H6 spectral setup). In addition, the maximum number of PRD iterations per main iteration is set to 10 and the convergence limit of PRD iterations in each main iteration is 1.0$\times10^{-2}$. We also created a wavelength table from 10 to 55000 $\AA$ in order to calculate $F_{\nu}$ for most frequencies. For further analysis, we also included C~I, C~II, and C~III atoms in LTE as background opacities, in addition to the H6 spectral setup's species, which we refer to as the H6CIII spectral setup hereafter.

\subsection{YMDF model components}
\label{subsec:YMDFmodel}

We developed the YMDF model\footnote{https://github.com/cepylka/ymdf} to reproduce the time-dependent spectra of flaring young M dwarfs.
This module predicts spectral evolution of individual flares utilising publicly available radiative outputs from the F-CHROMA grid of RADYN stellar flare simulations \citep{2024ApJ...969..121K,2022FrASS...934458K}. Those build upon foundational work by \citet{1992ApJ...397L..59C, 1995ApJ...440L..29C,2015ApJ...809..104A} and \citet{2023A&A...673A.150C}.

To calculate the flux integrated over the TESS wavelength range of the early M dwarfs in our sample, we use the linear two-component model \citep{2022FrASS...934458K,2024ApJ...969..121K,2025ApJ...978...81K}:
\begin{equation}\label{eq:2}
{f}_{\mathrm{obs}}^{{\prime} }=\frac{{R}_{\mathrm{\star}}^{2}}{{d}^{2}} (\hat{X}_{1}{F}_{1}+\hat{X}_{2}{F}_{2}),
\end{equation}
where R$_\star$  is the radius of the flaring star observed by TESS, and d is the star distance from the Gaia astrometric space mission recent data release (GAIA DR3, see \citealt{2023A&A...674A..28F}). The fluxes $F_1$ and $F_2$ are radiative surface fluxes derived by the corresponding models with the pre-flare flux subtracted. $\hat{X}_{1}$ and $\hat{X}_{2}$ give the best-fit filling factors of both RHD model components, representing the contribution of each model flux in the resulting flare flux.
These filling factors represent a physical area comprising multiple flare-related processes occurring simultaneously. A constant electron beam in the cF13-500-3 model and a pulsed electron beam in the m2F12-37-2.5 model treated as time-resolved phenomena with durations of several seconds. These components are averaged over a specific time interval in the models (0-9.8 s and 0-4.2 s for the m2F12-37-2.5 and cF13-500-3 models, respectively) to reflect multiple such events within the flare region, thereby capturing the spatio-temporal complexity of the flare evolution.  To obtain the spectra at the peak of the flare, we co-added time-resolved spectra from the RH 1.5D. For this, we run the RH 1.5D in the H6 and H6CIII spectral setups separately. From the flare flux at the peak, we obtained a temporal evolution of the flare employing several temporal flare models widely recognized in the literature, and we further discuss them in Sect.~\ref{subsec:sydi}.

Due to the linearity of the integration, the model enables two key applications: i) generating synthetic flux densities over extended wavelength ranges at the peak of the flare; ii) calculating integrated fluxes within specific spectral band passes (FUV and TESS).
We calculated $F_1$ and $F_2$ flux densities integrated over the TESS wavelength range (i.e., $\int R_{\mathrm{res}(\lambda)}{F}_{\lambda,1}d\lambda$ and $\int R_{\mathrm{res}}(\lambda){F}_{\lambda,2}d\lambda$, where $R_{\mathrm{res}}(\lambda)$ is the TESS response function) from m2F12-37-2.5 and cF13-500-3 atmosphere models using RH 1.5D (see for details \citetalias{2025A&A...700A..53M}, Eq.~3). To obtain the best fit of $\hat{X}_{1}$ and $\hat{X}_{2}$, we performed a systematic parameter search for $\hat{X}_{1}$ and $\hat{X}_{2}$ using Markov Chain Monte Carlo (MCMC) sampling with the `emcee' package \citep{2013PASP..125..306F}. For FUV spectral fitting, we retain the established methodology while omitting instrumental response corrections (see \citetalias{2025A&A...700A..53M}, Appendix A.5).

\subsection{Synthetic distributions}
\label{subsec:sydi}
Given that the equivalent duration (ED) serves as a proxy for flare energy, the derived model coefficients can be linked to ED values, thereby enabling amplitude estimation for flares of specified equivalent durations. ED quantifies flare energy relative to the star's quiescent luminosity through the integral relationship:
\begin{equation}\label{eq:7}
ED = \int \left( \frac{F_{\mathrm{flare}}(t)}{F_{\mathrm{quiescent}}} - 1 \right) dt \\[2ex]
\end{equation}
where  $F_\mathrm{quiescent}$ denotes the quiescent stellar flux and $F_\mathrm{flare}(t)$ represents the time-dependent flux during the flare event with duration $t$.  ED serves as a direct proxy for the specific flare energy when scaled by the star's quiescent luminosity in a given wavelength range.

The temporal evolution of the flare flux, $F_\mathrm{flare}(t)$, can be reconstructed using several established models from the literature. Our Python implementation includes three approaches: the temporal flare model described by \citet{2022AJ....164...17T}, the framework introduced by \citet{2014ApJ...797..122D}, and a modified Gaussian rise/exponential decay model based on the work of \citet{2024AJ....168...60F}. Hereafter, we refer to these as the Mendoza, Davenport, and Feinstein models, respectively.

These single-flare time-resolved models require three principal parameters: amplitude, duration, and peak times. We first implement the single-flare model with the amplitude equal to unity, and arbitrarily choose the peak time. We sample duration randomly from the distribution explained in \citet{2023A&A...669A..15Y} and \citet{2024A&A...689A.103Z}. They identified a distinct power-law correlation between flare duration $D$ and released energy $E$, characterised by $D \propto E^{\gamma}$, for events exceeding 10-minute durations. Below this temporal threshold, the relationship bifurcates, revealing that short-duration flares ($D < 10\,\text{min}$) can exhibit energy release comparable to their longer-duration counterparts. For the M stars, the power-law index was constrained to $\gamma$=0.3 \citep[][Fig.\,14]{2023A&A...669A..15Y}. The resulting normalised flux can be scaled by the peak flux amplitude derived from the YMDF model. By adding the quiescent stellar flux, the complete temporal evolution of the surface flux is produced. Integrating the model spectra over a specified wavelength range allows fitting of the resulting light curve to observations, such as the TESS white-light curve or the light curve obtained by integrating over the HST-COS G130M grating wavelengths. Subsequently, the model spectra can be compared to spectrally resolved HST-COS FUV data at specific time points, such as pre-flare or flare peak.

In \citetalias{2025A&A...700A..53M}, we proposed that a piecewise power law provides a more accurate representation of observed flare energy and EDs distributions in young M dwarfs. The YMDF model implementation enables both sampling of EDs from broken power-law as well as single power-law distributions. For every flare in the simulated distribution, EDs are used to determine the coefficients of the m2F12-37-2.5 and cF13-500-3 atmosphere models, enabling the simulation of the flare's spectral-temporal profile. This results in complete synthetic spectra of a star evolving for prolonged periods of time. These spectra can subsequently serve as inputs for exoplanetary atmospheric studies, providing critical data for chemistry kinetic models across a broad wavelength range from 10 to 50,000 $\AA$.

 \section{Results}
\label{sec:results}

\subsection{Light curve fitting and analysis.}
\label{sec:resultfsp}

\begin{figure*}[h!]
\sidecaption
\includegraphics[width=12.5cm]{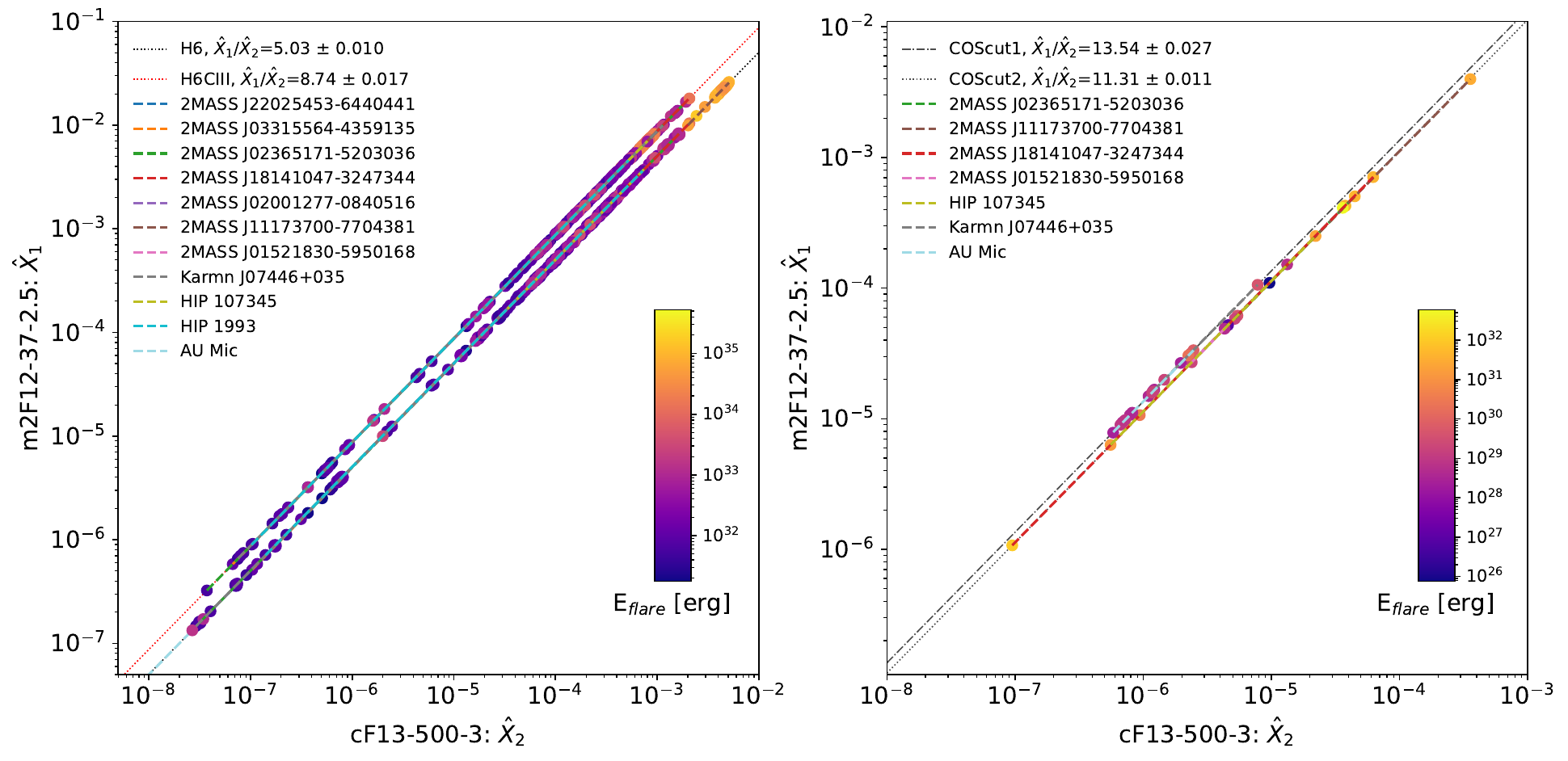}
\caption{Coefficients of the fitted models for flares in young M dwarfs. The values of $\hat{X}_1$ and $\hat{X}_2$ are plotted on the y- and x-axes, respectively. Each point represents an individual flare event and is colour-coded by the flare energy released in the TESS (left) or the FUV bandpass (right). Model fitting was performed using multiple setups: H6 and H6CIII for the TESS data, and COScut1 and COScut2 for the HST-COS data. Linear regressions fitted to the flare events for each star are shown as dashed coloured lines. To guide the eye, in the panels, the H6, H6CIII, COScut1 and COScut2 ratios are shown as red dotted, black dotted, grey dash-dotted and grey dotted lines, respectively.}

\label{fig:4}
\end{figure*}

We analysed the flare data in the TESS bandpass in our sample of 11 young M-K dwarfs by fitting the two-component model we described in Section \ref{subsec:steatmodel}. The TESS bandpass ($\sim$6000–10000 $\AA$) centred on the Cousins I-band, is largely devoid of prominent atomic and ionic emission lines, being dominated by the stellar continuum and a few broad molecular features \citep{2015JATIS...1a4003R}. This red-optical to near-infrared coverage was chosen for the TESS survey to maximise sensitivity to small planets transiting cool, red stars, but it lacks the rich array of diagnostic lines characteristic of the far-ultraviolet, resulting in a fundamentally different spectral information content. Fitting the simplified model based on the H6 spectral setup yields a ratio between the low energy m2F12-37-2.5 and the high energy cF13-500-3 stellar atmosphere models of R=5.03$\pm$0.01. Both coefficients, $\hat{X}_{1}$ and $\hat{X}_{2}$, from Eq.~\ref{eq:2} were free parameters while using MCMC sampling to fit to the observed data. Figure \ref{fig:4} presents the results of fitting $\hat{X}_{1}$ and $\hat{X}_{2}$ to observational data from TESS (left panel).
We further tested the sensitivity of these wavelength ranges to the nomenclature of lines, using the H6CIII spectral setup in the RH 1.5D calculations. This inclusion shift the mean ratio to R=8.74$\pm$0.02 for TESS range flares.

We proceed to analyse flare data in FUV in the same way as for TESS data. The 1040--1350 $\AA$ wavelength range is densely populated with spectral lines, making it a particularly rich region for stellar diagnostics. Among the most significant features in the FUV interval are strong emission lines (C II, C III, Si III, Si IV, N V), and the prominent and ubiquitous hydrogen Ly-$\alpha$ line at 1215.67 $\AA$ \citep{2017ARA&A..55..159L, 2013ApJ...763..149F}. Fitting the model based on the simplified H6 spectral setup is therefore a challenging task. To address these difficulties, we implemented the Kurucz line list \citep{1995KurCD..23.....K} in the RH 1.5D calculations in the range of 1083.990 - 1359.275 $\AA$ for species stated in \citet[Table~A1]{2022AJ....164..110F} and the H6 spectral setup species. Additionally, we adjusted the wavelength window for integration of the flux density resulting in two regimes. The first covers $\lambda \in$ [1060, 1360] $\AA$, excluding the interval $\lambda \in$ [1210, 1225]~$\AA$ to mask the Ly$\alpha$ emission. The second spans $\lambda \in$ [1170, 1430]~$\AA$. The HST-COS G130M grating configurations are discussed further in Appendix \ref{sec:appendixcos}. Hereafter, we refer to these spectral setups as COScut1 and COScut2, respectively. We used the time-tagged spectra (the detailed methodology can be found in \citetalias{2025A&A...700A..53M}) and integrated flux in FUV-B and FUV-A sectors for every timestamp. In these analyses, we produced light curves with 10 sec cadence to reach better precision in the fitting. As before, both coefficients, $\hat{X}_1$ and $\hat{X}_2$. in Eq.~\ref{eq:2} were treated as free parameters in the fitting while performing the MCMC sampling. We plotted them in Fig.~\ref{fig:4} (right panel) for the stars with a flare count $\ge$2 in FUV and the corresponding spectral setups.

For each star in the sample, we determine the best-fit coefficients, using TESS photometry data of flare events with ED>15 s and all available data in FUV. The number of flares used in the analyses is stated in Table \ref{tab:1}. The ratio between the fitted coefficients is indicated in each panel of Fig.~\ref{fig:4} for all stars in the sample, and, for the corresponding spectral setups, the mean values of the ratios $\hat{X}_1$/$\hat{X}_2$ are included in Table \ref{tab:2}.

While the component ratio is sensitive to the spectral setup, our further experimenting with incorporating additional emission lines in the COScut1 and COScut2 setups moves the initial ratio values R$\sim$11-13 closer to R$\sim$7-8, consistent with TESS results and prior NUV analyses (see, e.g., \citealt{2025ApJ...978...81K}). While including an exhaustive spectral line list in the radiative transfer RH1.5 calculations should in principle produce more reliable results, such calculations are computationally demanding and often suffer convergence issues. Overall, while the H6 model may be considered oversimplified, and the H6CIII model presents a ratio more consistent with previous studies, both models yield satisfactory fit to the time-integrated flux exhibited by flares in TESS and FUV light curves, as we demonstrate below in Sect.~\ref{sec:resultteif}.

\begin{table}
\scriptsize
\caption{\label{tab:2}Ratio of the model coefficients $\hat{{X}}_{{1}}$/$\hat{{X}}_{{2}}$.}
\centering
\resizebox{0.47\textwidth}{!}{\begin{tabularx}{9cm}{l c c c } \multicolumn{1}{r}{} & TESS & COScut1 & COScut2 \TBstrut\\
\hline
\multicolumn{1}{l}{Spectral setup /$\lambda$ } & [6000;10000] $\AA$ & [1060;1210]$\&$[1225;1360] $\AA$ & [1170;1430] $\AA$  \TBstrut\\
\hline
H6 & 5.03$\pm$0.010 & - & - \TBstrut\\
H6CIII& 8.74$\pm$0.01 & - & -  \TBstrut\\
H6COS& - &  13.54$\pm$0.027 & 11.30$\pm$0.014 \TBstrut\\
\end{tabularx}}

   \tablefoot{In RH 1.5D calculations, the H6 and H6CIII spectral setups are used with the wide range of $\lambda \in$ [10;55000] $\AA$. The H6COS setup is used with the $\lambda \in$ [220;40500] $\AA$ range. }
\end{table}

\subsection{Temporal evolution of individual flares}
\label{sec:resultteif}

Previously, we used bulk flare data to fit model coefficients and determine the median ratio between low- and high-energy  electron beam models. We now focus on individual flare light curves of different energies, temporal evolution and duration, observed in TESS and FUV for stars in our sample. The morphology found in the observed flares vary largely, both in the TESS \citep{2014ApJ...797..122D,2025ApJ...984..186Y} and FUV \citep{2018ApJ...867...70L,2018ApJ...867...71L,2019ApJ...871L..26F,2025AJ....170..249D} bandpasses. We plotted the temporal evolution of 16 flares in our sample in Fig.~\ref{fig:87} and the observational data for these flares are stated in Table~\ref{tab:102} for TESS and Table~\ref{tab:101} for FUV flares.

We further selected several observed flares and compare their fits with the flux obtained from models using inverse calculations. First, we determined the coefficients of the model components m2F12-37-2.5 and cF13-500-3 by utilizing the observed ED and the ratios derived from our model setups. This involves integrating the contributions of each component over the corresponding wavelength ranges for TESS/FUV ($F_1$ and $F_2$), as well as integrating the preflare flux $F_0$ within the model framework, and substituting these values into the ED equation (see Eq.~\ref{eq:7}, full derivation can be found in Appendix~\ref{sec:appendixYMDFED}). Second, the resulting coefficients allowed calculation of the peak flare flux (Eq.~\ref{eq:2}). Third, we employed a temporal model (the Mendoza model here) to scale the peak flux across all time stamps, thus capturing the temporal evolution of the spectrum. Throughout this work, we refer these temporally and spectrally-resolved model outputs as the "inverse models".

\begin{figure*}\resizebox{\hsize}{!}{
   \centering
\includegraphics{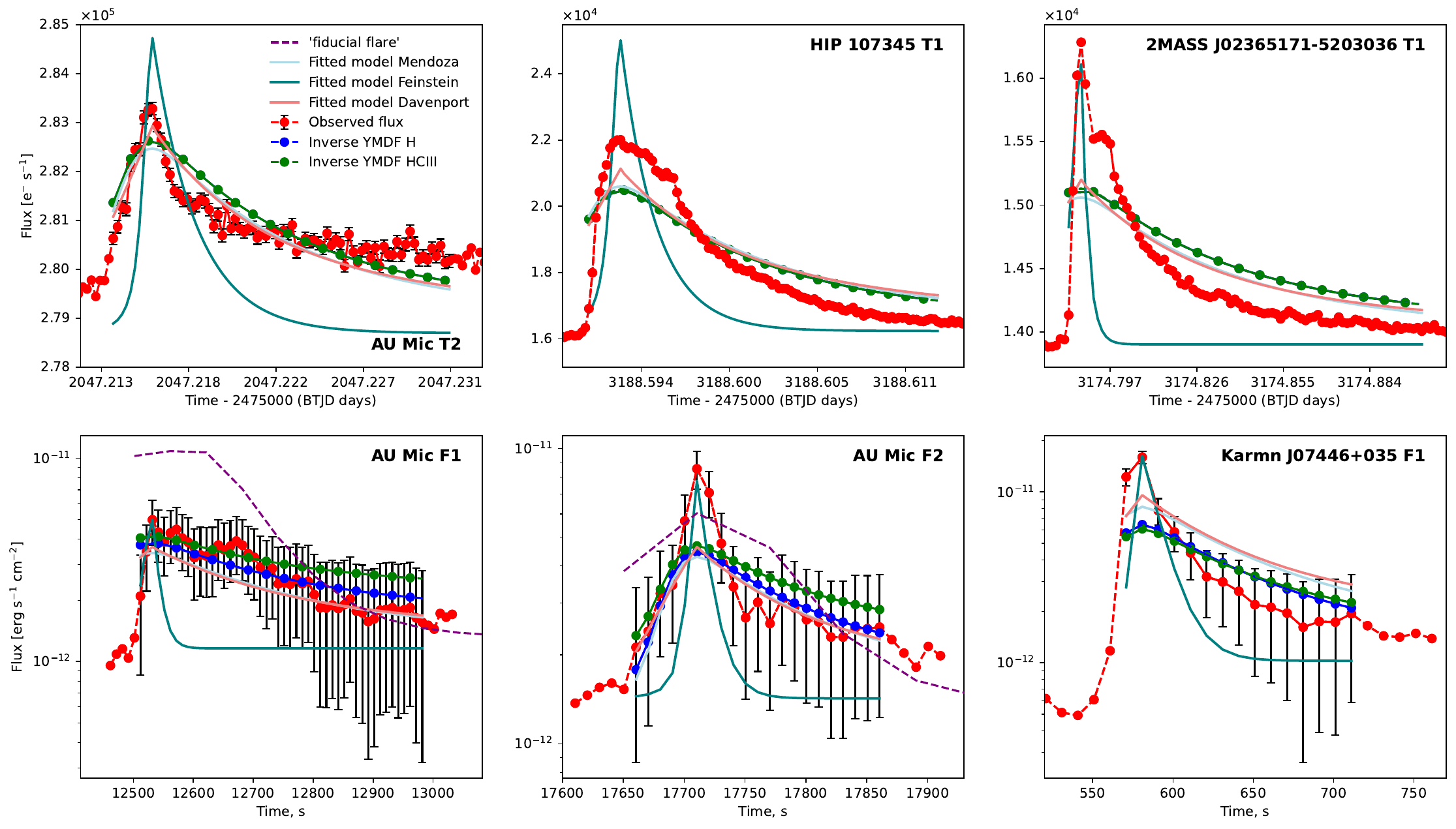}
}
\caption{The light curves of several flares observed in TESS and HST-COS. The upper row shows TESS flux light curves (e$^-$ s${^{-1}}$) for flares denoted as AU~Mic T2, HIP~107345 T1 and 2MASS~J02365171-5203036 T1 (the observation details are stated in Table \ref{tab:102} in Appendix \ref{sec:appendixTBF}) as dashed lines with red spheres for 20-second observation stamps and black error bars with capped ends. Note that errors can be smaller than the sphere size. Three temporal flare model MCMC fitting results are shown: Mendoza (light blue), Davenport (light coral), and Feinstein (teal) models fitted to the observed data. The H6 and H6CIII models, that were inversely calculated using observed EDs for these particular flares shown as solid blue and green lines with data-point spheres. In the upper row, the H6 model is hidden by the H6CIII due to overlap as their flux values are very similar. The lower row shows FUV HST-COS flux observations for flares denoted as AU~Mic F1, AU~Mic F2, and Karmn~J07446+035 F1, plotted with consistent styles and model fits as the upper row. The details of observations for these flares can be found in Table \ref{tab:101} in Appendix \ref{sec:appendixcos}.
}
\label{fig:85}
\end{figure*}

Results for three TESS flares and three FUV flares are shown in Fig.~\ref{fig:85} in the upper and lower rows, respectively. Observed data (red spheres with black error bars) are plotted along with the temporal flare models, illustrated here by coloured lines: Mendoza (light blue), Davenport (light coral), and Feinstein (teal). Inverse models of the H6 and H6CIII setups, derived using EDs from observed data, are shown as solid blue and green lines with circle markers, respectively. It is evident that the temporal evolution of flares in our sample is not fully captured by these models: the Mendoza model, characterized by rounded peaks and slow rise and decay, underperforms; the Davenport model exhibits a somewhat sharper peak but still fails to reproduce the full amplitude; and the Feinstein model reproduces peak amplitudes in many cases but inadequately models the slow decay. However, the simplified models can closely reproduce the emission observed by TESS and by COS-HST. We calculated time-integrated fluxes in HST-COS FUV (erg cm$^{-2}$) and TESS (electron counts), reporting these values for all models and observed data in Table~\ref{tab:10} (TESS) and Table~\ref{tab:11} (FUV). We found that the Mendoza and Davenport models produced time-integrated fluxes similar to the observed values, whereas the Feinstein model underestimated this value. The simplified inverse models H6 and H6CIII yielded results comparable to those obtained by directly fitting coefficients to observed data and, in some cases, outperformed the Mendoza model, which most closely reproduces the time-integrated observed flux, in the entire dataset.

\subsection{Model comparison to spectrally resolved data.}
\label{sec:resultmis}

RHD models of chromospheres predict the formation of a hot, dense condensation layer as a result of electron beam heating. When the energy flux of the electron beam is sufficiently high, the flaring regions reach continuum optical depths of approximately $\tau$$\sim$1, producing spectra characterised by temperatures around T$\sim$10$^4$ K and exhibiting a small Balmer jump. These features are consistent with the optical flare signatures observed in dMe stars (e.g., \citealt{1992ApJS...78..565H,2013ApJS..207...15K,2016ApJ...820...95K}). \citet{2019ApJ...871L..26F} employed these parameterisations to explore the modifications required in RHD models to reproduce the hot, blue FUV continuum emission detected during the GJ 674 flare. While RHD models with very high electron beam energy fluxes (10$^{13}$ erg cm$^{-2}$ s$^{-1}$) can generate the $\sim$10$^4$ K thermal component observed in previous flares, they are unable to account for the blue FUV continuum, as the model spectra turn over in NUV \citep{2015SoPh..290.3487K}.
\begin{figure*}
\includegraphics[width=18cm]{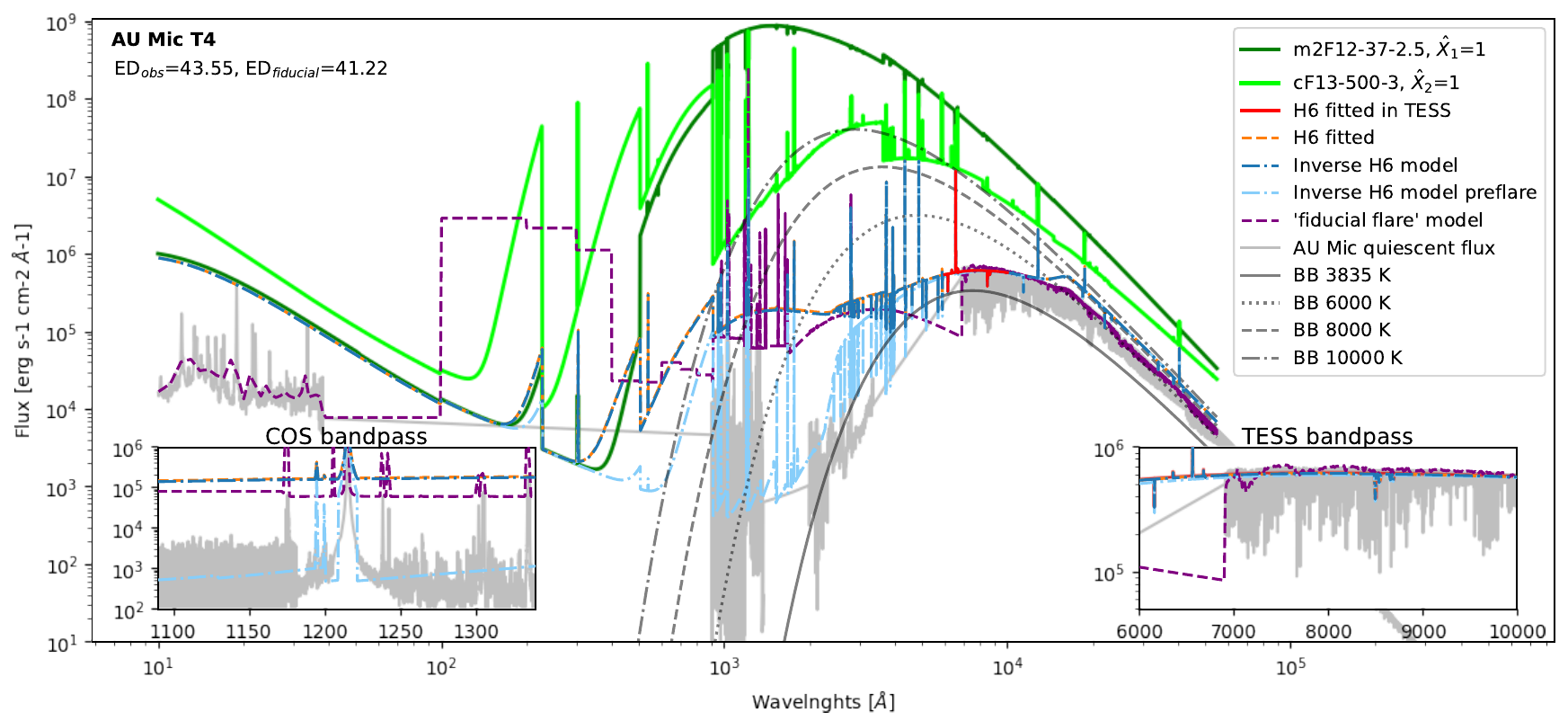}
\caption{Surface flux density observed and modelled. The m2F12-37-2.5 (dark green) and cF13-500-3 (light green) stellar atmosphere models are plotted at coefficients $\hat{{X}}_{{1}}$=1 and $\hat{{X}}_{{2}}$ = 1 (the theoretical assumption of the entire stellar surface exhibiting flaring activity). For AU Mic during a flare, the red solid and orange dashed lines show the fitted model spectrum at flare maximum using coefficients from the example fit (the flare data are stated in Table \ref{tab:102} in Appendix \ref{sec:appendixTBF}). The inverse H6 model at flare maximum, pre-flare and the `fiducial flare' model from \citet{2018ApJ...867...71L} with a comparable ED is plotted as blue dash-dotted, light blue dashed, and purple dashed lines, respectively. Blackbody curves for 3850 K, 6000 K, 8000 K, and 10,000 K are included in black solid, dotted, dash-dotted and dashed lines. The panchromatic AU Mic flux from \citet{2022AJ....164..110F} is plotted in light grey. Insets highlight the FUV (lower left) and TESS (lower right) wavelength ranges.}
\label{fig:6}
\end{figure*}

In Fig.~\ref{fig:6}, we plotted a comparison of the surface flux as a function of the wavelengths at the peak of the flare, highlighting the performance of the two stellar atmosphere models. The m2F12-37-2.5 and cF13-500-3 models, shown as dark and light green lines, respectively, were plotted assuming that the entire stellar surface is flaring. Although this scenario is unphysical, it enables a direct visual comparison of the spectral energy distributions predicted by each model across a broad wavelength range. We plotted here one example of the resulting spectra at the peak of the flare for AU Mic. The fitting procedure was performed for a median size flare, yielding ED$_\mathrm{model}$=43.55 s and a total energy release of 4.13$\times$10$^{33}$\,erg in the TESS bandpass. The corresponding fit is plotted as the red solid line. The obtained coefficients are used to produce the model spectrum at the peak of the flare in the H6 setup covering the entire model range ($\lambda \in$ [10;55000] $\AA$, shown as the orange dashed line). We determined $\hat{X}_1$ and $\hat{X}_2$ of the inverse H6 model from observed ED and the R=5.03. It is shown as a blue dash-dotted line, with the pre-flare state as a light blue dashed line. Blackbody spectral models with temperatures of 3850 K, 6000 K, 8000 K, and 10,000 K are included for reference. The panchromatic flux of AU Mic \citep{2022AJ....164..110F} is shown in light grey, empirically benchmarking for the model predictions. The additional panels focus on the FUV and TESS wavelength ranges, where the differences in model components' behaviour and the correspondence to the observed fluxes are most pronounced. For this medium-sized flare, the right inset (TESS bandpass) shows a modest increase in the continuum, whereas the left inset (FUV bandpass) reveals a pronounced continuum enhancement.

We further analysed the spectra of the sample stars by fitting the derived model-dependent parameters for several spectral setups.  \citet{2022AJ....164..110F} identified the continuum regions in the AU Mic spectrum, specifying these areas as being free of emission features. We adopted this approach to extract these regions from each time-tagged observed flux, subsequently rebinning the flux into 0.05 $\AA$ bins and further smoothing the signal by applying a rolling average over every 20 values, reaching an effective resolution of 1 $\AA$. We experimented with various methods of binning data and found this to be the optimal approach for effectively revealing the continuum. The H6 and H6CIII spectral setups have a native resolution of 1-10 $\AA$, while the COScut setups native resolution is 0.5-20 $\AA$. All setups treat key spectral lines with enhanced resolution; therefore, the applied binning and smoothing provide an effective intermediate resolution.

\begin{figure*}[h!]
\sidecaption
\includegraphics[width=12.5cm]{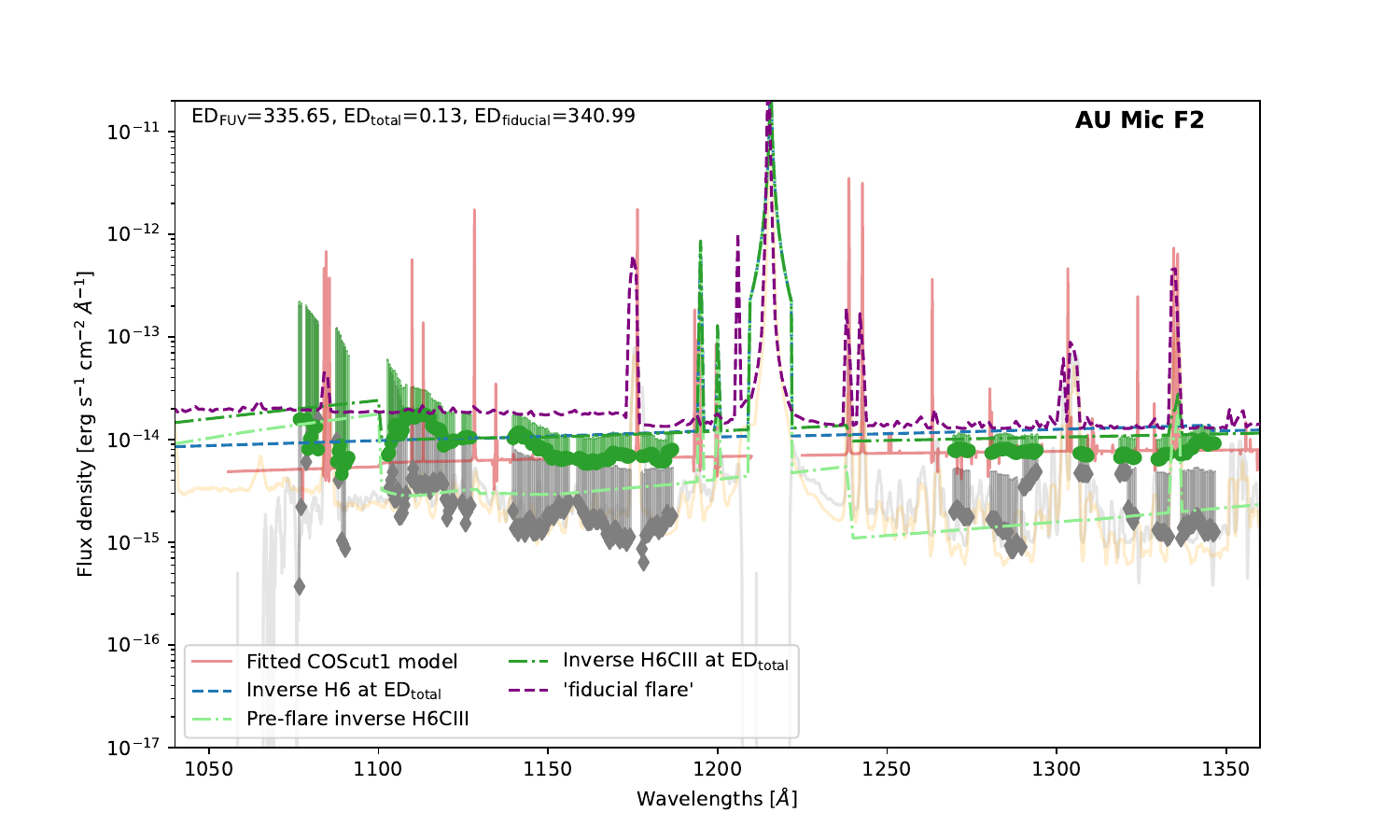}
\caption{Flare spectra in FUV HST-COS. Green circles and grey diamonds indicate the continuum flux at the flare peak and preflare, respectively, with uncertainties displayed as corresponding light-coloured lines; only the upper error bars are shown for visibility, as the errors are symmetric. The best-fit model to the continuum fluxes, derived using the COScut1 spectral setup, is plotted as a light red line. The blue, green, and light green dashed lines correspond to the H6 and H6CIII setups at the peak, and pre-flare H6, respectively. The light grey, light orange and purple dashed lines represent: the sum of the extracted one-dimensional spectra from all exposures in the orbit, the smoothed panchromatic spectrum of AU Mic from \citet{2022AJ....164..110F}, and the `fiducial flare' model from \citet{2018ApJ...867...71L} at a similar ED, respectively.}
\label{fig:15}
\end{figure*}

In Fig.~\ref{fig:15}, we compare the obtained continuum spectra at the peak of the FUV flare in AU Mic. In order to extract the observed continuum flux, we rebinned the flux at the timestamp of 17660.5 s from the start of observation for the flare peak and at the timestamp of 16700.0 s for the quiescent continuum (plotted here as green spheres and grey diamonds, respectively).
To interpret the continuum enhancement, we plotted the fitted COScut1 setup at the flare peak. For additional context, we included comparisons with the inverse models H6 with R=5.03 and H6CIII with R=8.74 calculated from the observed flare ED.

We note that, the COScut1, H6 and H6CIII setups all results in a good-to-eye fit in FUV-A segment  (1225-1360~$\AA$) and partially in FUV-B (1130-1210~$\AA$), indicating that these models can potentially reproduce the continuum rise during the flare. However, the continuum rise observed in the blue short-wavelength region ($\sim$1110-1120$\AA$ ) of the FUV-B segment remains unaccounted, and the rise in the H6CIII peak and preflare models is shifted to wavelengths shorter that $\sim$1110 $\AA$. This rise observed in the blue FUV-B continuum has been detected and analysed in detail by \citet{2022AJ....164..110F}. We continue this discussion in Sect.~\ref{sec:discsdc}.

Our attempt to fit the preflare flux to the observed FUV data was unsuccessful for two stars, both of which exhibit comparatively weak flare events, making accurate fitting particularly challenging. First, in 2MASS~J18141047-3247344 (HD 319139), a spectroscopic binary consisting of K5 and K7 dwarfs \citep{2012MNRAS.425..950N}, the YMDF H6 model underestimated the preflare quiescent flux by two orders of magnitude. This discrepancy may be attributed to the generally lower levels of magnetic activity observed in K dwarfs compared to M dwarfs, as well as the complex nature of the binary system; the fit fails as expected since the model assumes a single star, while in reality, the system consists of two unresolved stars. In 2MASS~J11173700-7704381, the youngest star in our sample at $\sim$5 Myr old, we observed a similar issue, with the YMDF model significantly underestimating the preflare quiescent flux. To better understand how the model could replicate the flares in a range of M-K dwarfs, additional observations in the FUV and NUV are needed. Such data would help to more accurately characterise the preflare emission and flare properties of young, low-mass stars, providing further constraints for theoretical models.

Interestingly, for Karmn~J07446+035 (YZ CMi), we achieved a successful fit in FUV-A (see Fig.~\ref{fig:26}, left panel, in Appendix \ref{sec:appendix1}), despite it being the dimmest and oldest star in our sample. Similarly to AU Mic, in YZ CMi, we observe a continuum rise in the blue end of the FUV-B segment (approximately $\ge$1130 $\AA$) comparable to features reported in AU Mic flares \citep{2022AJ....164..110F,2002ApJ...581..626R}, as we discuss further in Sect.~\ref{sec:discsdc}.
YZ CMi, M4.5 dwarf, is approximately 50 Myr old, and is thought to be near the transition line between partially and fully convective regimes (see, e.g., \citealt{1997A&A...327.1039C,2000ARA&A..38..337C}). Our model may slightly overestimate the preflare quiescent flux, but overall, the agreement is good. YZ CMi is well known for its high level of magnetic activity, with flares observed across multiple wavelength regimes over many years \citep{1961ApJ...133..914R,2017ApJ...837..125K,2023ApJ...945...61N}. This extensive observational history makes it a valuable benchmark for testing flare models in active M dwarfs. These results indicate that the YMDF model, though presently tested on young early M dwarfs, may also be applicable to stars with properties beyond those in our current sample.

\subsection{FFDs in simulated populations.}
\label{sec:resultffd}

In order to simulate star flare activity, the YMDF module: i) simulates the ED distribution based on a single or piece-wise power law; ii) produces randomly chosen durations from a bifurcated distribution \citep{2023A&A...669A..15Y,2024A&A...689A.103Z} based on given ED; iii) determines flare start times according to the probability of event occurrence described by a Poisson distribution; iv) injects obtained events at corresponding time stamps according to chosen cadence to the quiescent spectra resulting temporally and spectrally resolved dataset, which represent flare activity for a chosen period.

For practical application to flare frequency analysis, we chose the period of 51 days to facilitate direct comparison with the approximate duration of the observation of AU Mic in TESS sectors 1 and 27 with excluded gaps. Thus we generated synthetic light curves and spectra compatible with observed flare frequency distributions (FFDs). The model's output includes time-resolved spectra spanning the 10–50,000 $\AA$ range with a temporal resolution of up to 20 sec cadence, sufficient to resolve the impulsive phases of M dwarf flares, as \citet{2014ApJ...797..121H} similarly demonstrated for Kepler data with a 60 sec cadence.
\begin{figure}
\includegraphics[width=8.5cm]{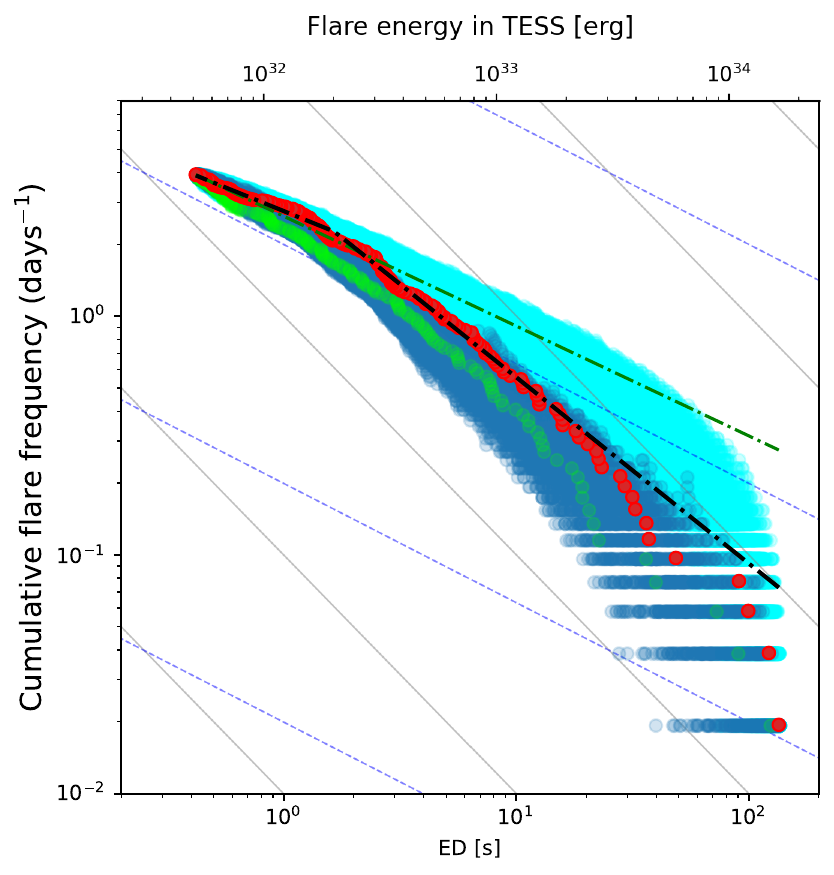}
\caption{Cumulative FFDs (scatter) of simulated ED distributions for broken power law relation (blue spheres) with $\alpha_1$=1.39 and  $\alpha_2$=1.8 found from the observed distribution of AU Mic white light flares. The intrinsic observed AU Mic's FFD is plotted as red spheres (adopted from \citetalias[Fig.~5, left panel]{2025A&A...700A..53M}). Simulated ED distributions for single power laws with $\alpha$=1.39 and $\alpha$=1.8 as light blue and lime spheres, respectively. The resulting distributions are from one example run with 1000 random samples, only plotted if  the sample meets p>0.01 in the K-S test. The grey and blue dashed guides correspond to power law coefficients $\alpha$=2.0 and $\alpha$=1.5, respectively, for a range of $\beta$ coefficients.}
\label{fig:5}
\end{figure}

We proceed to analyse the ED distributions in the synthetic stellar activity datasets generated by the YMDF module described above. We performed two complementary statistical tests to evaluate how well our synthetic population represents the observed data. The first is a Kolmogorov–Smirnov (K–S) test, which assesses the null hypothesis that both samples are drawn from the same parent distribution. The second is the Pearson $R$ test \citep{pearson1905general}, which measures the strength of the underlying linear correlation between the variables. Considered together, the results of these tests provide a quantitative basis for identifying the approach that offers the best agreement with the observations.

In Fig.~\ref{fig:5}, we present the results of an example of three simulation runs, each producing 1000 ED distributions, and examine the corresponding FFDs. As discussed in \citetalias{2025A&A...700A..53M} (i.e., Fig.~5, left panel), we employed a broken power law with the empirically derived slopes $\alpha_1$=1.39 and $\alpha_2$=1.8. In addition, we modelled the single power law populations based on $\alpha_1$ and $\alpha_2$, respectively. We applied the K–S test to assess whether the simulated and observed distributions originate from the same population, and only included distributions with $p > 0.01$ in the K–S test in the figure. To assess the robustness of our approach, we repeated the test 100 times. The results across runs were consistent: typically, up to 900 populations that successfully met our condition were generated using a single power law ($\alpha_1$), while approximately 200 -- 300 populations were generated using the broken power law ($\alpha_1$ and $\alpha_2$ ), and at most 2 populations with $\alpha_2$ alone pass the K–S test.

\begin{figure}
\includegraphics[width=9cm]{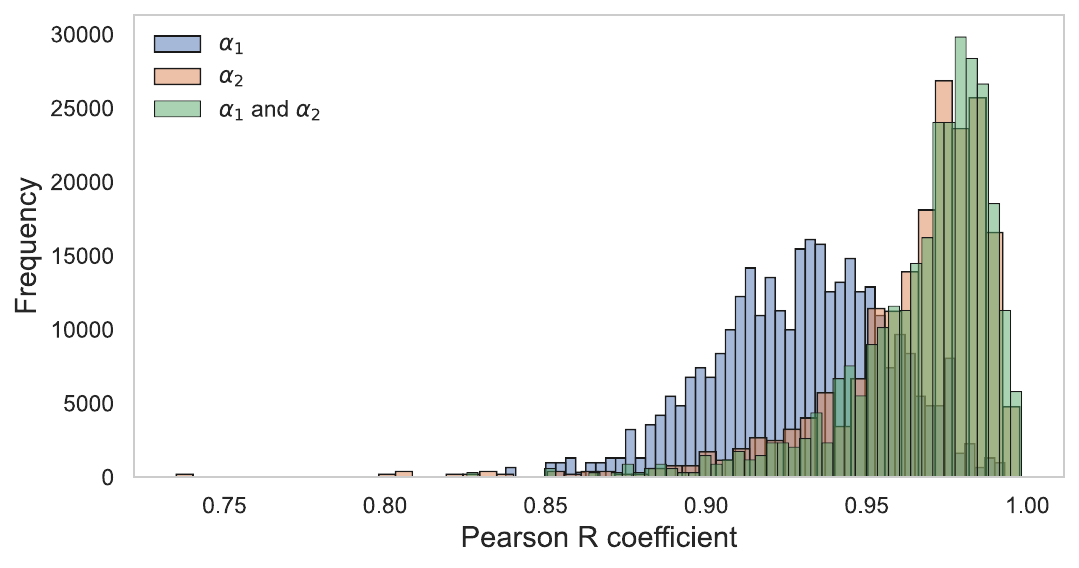}
\caption{Pearson $R$ coefficients indicating the linear correlation with the intrinsic observed AU Mic ED distribution: shown for simulated ED distributions generated using a broken power law with observed slopes $\alpha_1$=1.39 and $\alpha_2$=1.8 (green bars), and for single power law distributions using $\alpha_1$ (blue bars) and $\alpha_2$ (red bars), respectively. Results are based on an example run with 1000 random samples.}

\label{fig:61}
\end{figure}

We evaluated the linear correlation between the intrinsic stellar ED distribution and the simulated distributions by calculating the Pearson $R$ coefficients. The results are shown in Fig.~\ref{fig:61}, indicating that both the broken power law and the single power law with $\alpha_2$ provide samples that correlate with the observed intrinsic distribution better than the  distributions govern by the single power law with $\alpha_1$. For AU Mic, the peak of the distribution with $\alpha_1$ is shifted further from unity compared to the other two distributions. We analysed all stars in the sample using the same methodology and found our conclusions to be consistent throughout, likely due to the homogeneity of the young star sample. For some stars, the peak of the Pearson $R$ distribution with $\alpha_1$ can be as low as 0.6.

When combined, these results indicate that only the broken power-law approach meets the criteria of both statistical tests and provides the closest approximation to the observed distributions. Thus, by employing the YMDF model, we reconfirm the conclusions drawn in \citetalias{2025A&A...700A..53M}.

\section{Discussion}
\label{sec:discussion}
\subsection{Comparison of stellar activity models}
\label{sec:discsdc}
\citet{2018ApJ...867...71L} made an impactful contribution to the study of M dwarf activity by introducing the `fiducial flare'  model \citep{2022ascl.soft02012L}. Utilising FUV observations of several M dwarfs, they constructed an idealised flare template that captures both the spectral and temporal characteristics of typical stellar flares observed in these stars. This model is constructed from high-cadence FUV observations and is characterised by a box-car form, with rapid rise, long plateau and slower decay in flux. Spectrally, the model includes both continuum and emission line enhancements, reflecting the observed increases in chromospheric and transition region lines such as C II, Si III, and Si IV during flares. By quantifying these temporal and spectral components, the `fiducial flare' model provides a reproducible template for assessing the atmospheric response of exoplanets to stellar flaring, facilitating systematic studies of M dwarf activity (see, e.g. \citealt{2023MNRAS.521.3333L,2025arXiv250313353D}). The `fiducial flare' model has since become a widely used tool to investigate the effects of both rare, high-energy and frequent, low-energy flares on the chemical and physical environments of planets orbiting M dwarfs.

Following the routines in \citet{2022ascl.soft02012L}, we generated a synthetic `fiducial flare' stellar activity dataset for AU Mic spanning 51 days. Generating such a dataset requires access to the panchromatic spectrum of the star in its quiescent state. Although several M dwarfs from the MUSCLES survey, with their high-resolution spectral energy distributions (SED), were used to refine the `fiducial flare' model, these stars are generally older than 800 Myr and significantly less active than those in our sample. For AU Mic, however, a panchromatic spectrum \citep{2022AJ....164..110F} is available and has been used throughout this study. This enabled a direct comparison between the YMDF and the `fiducial flare' models. For further context, in Fig.~\ref{fig:6}, which presents the comparison in the observed and modelled surface flux density, we plotted the `fiducial flare' model as a purple dashed line for a flare with a similar equivalent duration of the flare we used for fitting (ED$_\mathrm{TESS}$=41.22\,s and 43.55 s, respectively). While the YMDF and `fiducial flare' models broadly reproduce the observed TESS-band emission in agreement with each other, their predictions diverge in the FUV and NUV. Here, the `fiducial flare' model shows reduced energy release in these ranges compared to the YMDF model.

Previously, our range of models provided satisfactory fits to the FUV-A continuum; however, notable discrepancies remain in reproducing the continuum rise observed in the blue portion of the FUV-B segment. Thermal bremsstrahlung (TB) is a principal emission mechanism responsible for the soft X-ray emission observed in solar flares \citep{1996AdSpR..17d...9S,2019ApJ...881..161M}. \citet{2022AJ....164..110F} modelled the continuum rise during AU~Mic FUV flares using blackbody profiles spanning 9,000-11,000~K. They found that fits converged only for electron number densities unrepresentative of typical chromospheric conditions ($\sim$10$^{22}$~cm$^{-3}$). From COS imaging, they confirmed that count rate enhancement is restricted to the spectral trace region on the detector, supporting that it is an astrophysical feature, and they concluded that while TB alone cannot explain the FUV excess, additional emission processes are likely involved.

The TB emission during AU~Mic flares predicts significant continuum enhancement at extreme-UV wavelengths, exceeding the brightness of bound-bound emission lines and recombination continua by several orders of magnitude. However, no such enhancement was observed with EUVE during AU~Mic flares \citep{1993ApJ...414L..49C,1996ApJ...466..427M} nor in other M dwarf FUV flare observations \citep{2019ApJ...871L..26F,2018ApJ...867...70L}. In our sample, in Karmn~J07446+035 we also observe a continuum rise in the blue end of the FUV-B segment (see Fig.~\ref{fig:25}, left panel), while the rest of the sample's stars were not observed in this particular spectral region. Therefore, the precise physical origin of the FUV-B blue continuum rise necessitates further investigation to account for unresolved processes, possibly involving TB emission, and thus not yet incorporated into our models.

We fitted the `fiducial flare' model to the observed FUV continuum. This model also does not reproduce the sharp rise in FUV-B continuum, but, overall, in FUV tends to slightly overestimate the observed flux when compared to flares of similar ED$_\mathrm{FUV}$ and generally, in this model, the FUV-B segment is elevated comparing to FUV-A, as illustrated in Fig.~\ref{fig:15} (observed ED=335.65 s; ED$_\mathrm{fiducial}$=340.99 s). For smaller flares, the `fiducial flare' model matches the observed continuum at the flare peak more closely (see Fig.~\ref{fig:26} in Appendix~\ref{sec:appendix1}, right panel; observed ED=156.04 s, ED$_\mathrm{fiducial}$=154.92 s). In contrast, for the largest flare in our AU~Mic sample, the `fiducial flare' model predicts a flux that considerably overestimates the observations (see Fig.~\ref{fig:26} in Appendix~\ref{sec:appendix1}, left panel; observed ED=639.31 s, ED$_\mathrm{fiducial}$=638.01 s).

A major challenge in modelling spectra during the flare in the `fiducial flare' model is the requirement of a panchromatic spectra across the 10–50,000~$\AA$ wavelength range. The `fiducial flare' model inadequately addresses gaps in spectral data (see e.g., Fig.~\ref{fig:6}, lower right inset, 6000–7000~$\AA$). For AU~Mic, this limitation results in an underestimation of flare energy in the TESS bandpass, yielding a modelled FUV energy release smaller than expected (the figure’s right inset). Therefore, while assuming continuous panchromatic spectral coverage, the ED predicted by the `fiducial flare' model would be expected to exceed the observed values. Conversely, in the FUV domain, where the panchromatic AU Mic spectra has full coverage, at similar EDs to observed, the `fiducial flare' tends to overestimate both the flux density and the time-integrated flux (see Table~\ref{tab:101}) compared to observed values, thus affirming that the model overpredicts the observed flux density at certain EDs. Since FFDs are predominantly derived from statistically robust, extensive TESS datasets, these modelling limitations would contribute to an excess of small flares and a deficit of mid-size flares, which we explore further in Sect.~\ref{sec:discffds}.

\subsection{Temporal evolution of a flare in stellar activity models}
\label{sec:disctef}

The morphology of flare events exhibits significant variability, reflecting complex and poorly understood underlying physical processes \citep{2014ApJ...797..122D,2014PASP..126..398H}. In our sample, we observe complex flare morphologies across different spectral bands, as illustrated for 16 flares exhibiting energies in broad range (ED$_{TESS}\in$[11.8-437.5], ED$_{FUV}\in$[8.69 - 639.31]) in the TESS and HST-COS FUV observations (Fig.~\ref{fig:87} in Appendix~\ref{sec:appendixfit}). Various parametrizations have been employed to describe observed flare profiles, including composite Gaussian-exponential models \citep{2022AJ....164...17T}. The analyses have distinguished impulsive and gradual phases in white-light flares \citep{2013ApJS..207...15K} and identified quasi-periodic oscillations \citep{2015ApJ...813L...5P}. High-cadence observations have revealed flare peak rollovers, motivating continuous models without discontinuous breaks between rise and decay phases \citep{2016ApJ...820...95K,2022ApJ...926..204H}.

We further analysed the temporal evolution of individual flares by integrating the time-resolved model spectra over TESS and FUV (excluding the detector gap) wavelength ranges into corresponding light curves of AU Mic (see Fig.~\ref{fig:85}) at EDs in TESS and FUV similar to observed. The fit for TESS in AU Mic T2 panel reveals that the temporal evolution of the `fiducial flare' lies outside the panel axis (upper row, left panel) as at the similar ED the `fiducial flare' model produces fluxes $\sim$2.35-2.55$\times$10$^5$ e$^-$ s$^{-1}$ which is much less that the observed quiescent flux of 2.79$\times$10$^5$ e$^-$ s$^{-1}$. Hence, the `fiducial flare' is omitted in this panel to enhance visual clarity. We attribute the lower-than-quiescent flux to a deficiency in the `fiducial flare' spectrum within the 6000–7000~$\AA$ region and to the model's approach in handling an incomplete panchromatic spectrum (previously discussed in Sect.~\ref{sec:discsdc}), while applying scaling for the box-car temporal evolution. Additionally, the time-integrated flux value for the `fiducial flare' is much lower than observed (see Table~\ref{tab:10} in Appendix\ref{sec:appendixfit}). However, in the AU Mic F1 and AU Mic F2 panels, we successfully plotted the `fiducial flare' temporal evolutions considering the similar EDs to the observed. The box-car temporal form of the `fiducial flare' fails to reproduce the observed temporal evolution of these flares in detail, but also overestimates the time-integrated flux (see Table~\ref {tab:11} in Appendix\ref{sec:appendixfit}.

To our knowledge, the literature lacks extensive and comprehensive statistical studies that thoroughly characterize the detailed morphology of stellar flares to allow realistically reproduce not only frequency but also temporal profile variability in flares. Consequently, in modelling temporally resolved spectral variations of stars, it is necessary to adopt simplified approaches to maintain computational feasibility. However, the YMDF module currently allows the selection among the three aforementioned temporal flare models, and future development may include implementation of random model selection/combination to enhance flexibility.

\subsection{FFDs in synthetic datasets}
\label{sec:discffds}
\begin{figure*}
\centering
    \includegraphics[width=18cm]{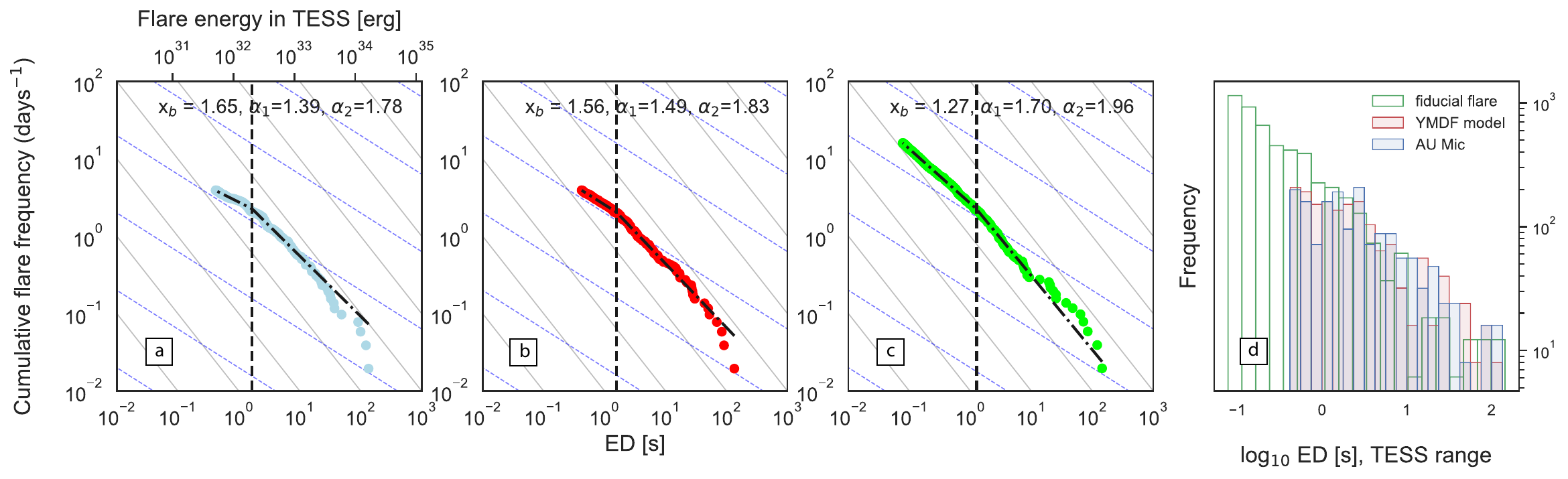}
\caption{Cumulative FFDs of simulated and observed EDs in the TESS range. The first three panels from the left show the corresponding FFDs in the TESS bandpass: observed data (panel "a"), a representative synthetic distribution generated by the YMDF model (panel "b"), and by the `fiducial flare' model (panel "c"). The synthetic distributions are produced for the period of observations corresponding to the observed periods for AU MIc in TESS ($\sim$ 51 days). Grey and blue dashed guides denote reference power-law slopes of $\alpha$=2.0 and $\alpha$=1.5, respectively. The right panel "d" shows the synthetic distributions, which are produced for the period of observations corresponding to the observed periods for AU MIc in TESS. Blue, red and green bars indicate distributions produced by the `fiducial flare' model, the YMDF model, and the observed ED distributions across two TESS sectors (1 and 27) for Au Mic, respectively.}
\label{fig:7}
\end{figure*}
\begin{figure*}
\sidecaption
\includegraphics[width=18cm]{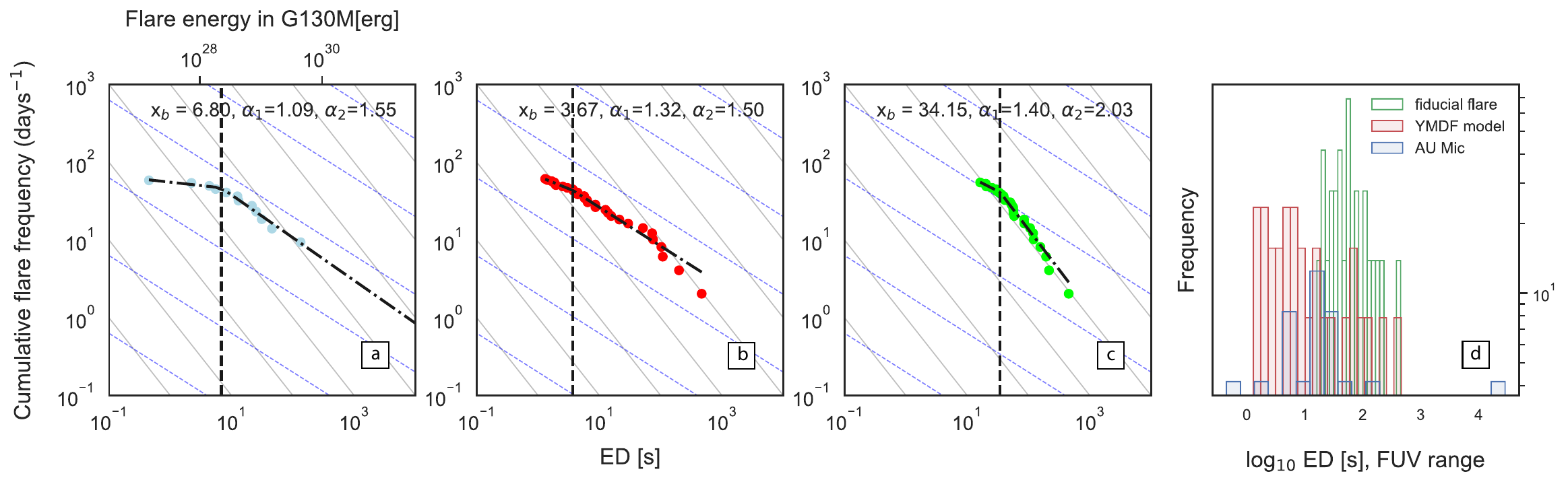}
\caption{Cumulative FFDs in the HST-COS FUV bandpass are shown in the first three panels from the left: observed data (panel "a"), a representative synthetic distribution from the YMDF model (panel "b"), and the distribution from the `fiducial flare' model (panel "c"). These synthetic distributions correspond to the AU Mic observation period in FUV ($\sim$ 11.7 hours).
The right panel "d" displays synthetic distributions produced over the AU Mic FUV observation period. The bars are coloured consistently with those in the preceding figure, except for corresponding to AU Mic observation period in FUV.}
\label{fig:8}
\end{figure*}

To assess the performance of flare models in reproducing observed flare activity, in Figure~\ref{fig:7}, we present cumulative FFDs of EDs in TESS for AU Mic, alongside simulations from the YMDF and ‘fiducial flare’ models. The YMDF model (panel b) closely matches observations (panel a), while the ‘fiducial flare’ model (panel c) exhibits discrepancies, notably overpredicting small flares and underpredicting medium-sized events. The panel (d) further compares ED distributions, demonstrating that a synthetic YMDF population following a broken power law with slopes $\alpha_1$ and $\alpha_2$ aligns well with the data. For the largest flares, however, both models are likely to yield similar results, accurately reflecting the rarity and energetic dominance of these events.
Extending the analysis to the FUV range (Fig.~\ref{fig:8}), the YMDF model exhibits an excess of small flares relative to observed FFDs, which can be attributed to use of the broken power law in the module with $\alpha_1$ and $\alpha_1$ equal to those observed. Nevertheless, it produces EDs closely matching the range of the observed values. In contrast, the `fiducial flare’ model shows a systematic shift toward higher EDs, indicating a tendency to produce more energetic flares than observed in FUV.

In \citetalias{2025A&A...700A..53M}, we found that the slopes of the broken power law differ between the FUV and TESS ranges; specifically, the high-energy slope ($\alpha_2$) in the FUV closely matches the low-energy slope ($\alpha_1$) in the TESS range. This correspondence,  also prominent in the synthetic ED distribution, suggests that a single broken power law, with $\alpha_1 \approx 1.5$ and $\alpha_2 \approx 2$, could adequately characterise the entire FFD, at least for young, active M stars with rotation periods shorter than 10 days. If supported by further observations, this unified description could offer a practical way to interpret flare activity across a wide range of energies in young, active stars.

These results support the proposition that young M dwarfs’ flare activity follows a unified broken power law, with FUV observations probing lower-energy flares below TESS’s detection threshold, and TESS capturing the larger, more energetic flares. Such a unified framework facilitates the interpretation of multi-wavelength flare activity and advances understanding of stellar magnetic phenomena in young active stars. Accurately capturing mid-sized flares is significant for exoplanet habitability through their influence on atmospheric chemistry and escape, and we argue that it is better achieved by the YMDF model.

\subsection{Observational and modelling challenges for M dwarf flare spectra}
\label{sec:disctef}

Required by previous models such as the `fiducial flare', the panchromatic spectra of M dwarfs remains challenging to obtain due to their intrinsic faintness, particularly in the UV and X-ray regimes, necessitating long exposure times with sensitive instruments. Coordinating simultaneous observations across the full spectral range is logistically complex and compounded by their high variability during flares, which renders non-simultaneous data unreliable for representing true spectral energy distributions (see, e.g., \citealt{2016ApJ...820...89F,2019ApJ...871L..26F}). However, the YMDF model has the potential to generate representative panchromatic spectra for stars lacking comprehensive observational data.

The YMDF model relies on the \citet{2024ApJ...969..121K} RHD model grid that demonstrated the ability to reproduce the behaviour of hydrogen lines, particularly in the NUV regime as shown in previous studies. \citet{2022FrASS...934458K} successfully modelled the impulsive phase of AD Leo’s Great Flare using RADYN simulations of electron beam heating, supporting similar approaches for fitting flare flux to the continuum while also reproducing observed hydrogen line profiles such as H$\gamma$. \citet{2025ApJ...982...98G} conducted a forward modelling of observed H$\alpha$ and H$\beta$ spectra from a small flare on an M4.5 dwarf using a comprehensive grid search across 43 RADYN models with electron fluxes ranging from 10$^{10}$ to 10$^{13}$ erg s$^{-1}$ cm$^{-2}$. Their results favour the less energetic mF11-17-3 model, consistent with the flare occurring on a less active, likely older and fainter M dwarf compared to our sample stars. Similarly, \citet{2025arXiv250906908N} compared observed H$\alpha$ and H$\beta$ line profiles from AU Mic flares with the aforementioned RHD model grid, finding that electron beam heating parameters reproducing NUV/optical continuum emissions also broadly reproduce the pronounced line broadening near flare peaks. The inferred electron beam flux densities of 10$^{12}$ to 10$^{13}$ erg cm$^{-2}$ s$^{-1}$ exceed those typical of solar flares, aligning well with our findings.

The most prominent line feature, the Balmer series, and particularly the Balmer jump at $\sim$3650\,\AA, shows that flare flux can exceed blackbody model predictions \citep{2013ApJS..207...15K,2019ApJ...871..167K}. Using a setup similar to ours, \citet[][Fig. 5]{2025ApJ...978...81K} demonstrate that, despite the lack of simultaneous optical spectra to constrain the Balmer jump and optical emission lines, the observed TESS flare-only flux supports the robustness of extrapolating the RHD model into the near-infrared.

From solar studies, elevated continuum emission extends into the NUV ($\sim$2000--3000\,\AA), indicating a common origin with the optical continuum \citep{2021A&A...654A..31J}. In the NUV, emission lines (notably Mg~II and Fe~II) contribute more significantly, making up 20-50\% of the NUV flux \citep{2007PASP..119...67H,2024MNRAS.533.1894J}, with higher-energy flares being more continuum-dominated. For example, \citet{2009AJ....138..633K} measured line contributions of $\sim$40\% in the 2510--2841\,\AA\ range for two GJ~1243 flares observed with HST-COS. However, our model setups, COScut1 and COScut2, incorporate all major lines as background opacities. Comparing these model setups with the H6 setup, both in light curves (Fig.~\ref{fig:85}) and spectra (Figs.~\ref{fig:15}, \ref{fig:26}, \ref{fig:25}), reveals comparable time-integrated fluxes, implying similarity in ED$_\mathrm{FUV}$, and the fitting of these models to the continuum fluxes is yielding similar results. This suggests that the simplified H6 approach is at least capable to adequately able to capture the flare energy budget, while addressing the challenge of modelling variable spectra in young M dwarfs.

The YMDF model in the H6 setup primarily captures the continuum enhancement observed during stellar flares, along with the prominent hydrogen lines evolution as shown in previous work. Multiple studies have demonstrated, that simple models that focus on the continuum rise and hydrogen recombination can provide a robust match to observed flare spectra, even when more detailed line emission modelling is not feasible. For example, \citet{1991ApJ...378..725H} found that roughly one-third of the line emission occurs during the impulsive phase, whereas the majority of the continuum emission is released during this same period. Because the continuum dominates the overall energy output, most of the total flare energy is also emitted during the impulsive phase, at least in the 1200–8000 $\AA$ wavelength range. It is worth noting that radiation emitted within this wavelength range is particularly relevant for studies of exoplanet atmospheres. Photons at these wavelengths drive the majority of photodissociation reactions \citep{2020ApJ...896..148R}, which can ultimately lead to atmospheric depletion.

Studies of M-star flares indicate that FUV emission can precede white-light emission, implying that it may originate from the initial heating, compression, and evaporation of plasma at the flare footpoints \citep{2019ApJ...871L..26F,2003ApJ...597..535H,2021ApJ...911L..25M}. Furthermore, flare observations of M dwarfs have reported temperatures reaching up to 20,000 K and in some cases as high as 40,000 K in both the optical and FUV regimes (see, e.g. \citealt{2019ApJ...881....9H,2019ApJ...871L..26F}). Although our model configuration is capable of reproducing the high energies released in both the FUV and NUV ranges, it is currently unable to replicate the observed differences in the onset times of white-light flares and FUV emission. This discrepancy highlights a limitation in the present modelling approach. Further investigation is required to identify and accurately parametrise the physical processes responsible for these timing differences.

\subsection{Summary and conclusions}
\label{sec:concl}

In this study, we tested several spectral setups of the combinations of the two RHD models with high- and low-energy electron beams. We found that the straightforward H6 non-LTE hydrogen atom model with the inclusion of several important atoms in LTE can satisfactorily reproduce the observed continuum rise in the TESS and partly in the FUV range for flares spanning a wide range of EDs. This approach models stellar flare continua for these two regions, and in NUV and optical according to previous literature (see, e.g. \citealt{2025ApJ...978...81K,2025arXiv250906908N,2025ApJ...982...98G}). Improvements should include expanding the atomic and molecular set, and enhancing the treatment of atmospheric structure and time-dependent effects. Comparison with high-cadence, multi-wavelength data will further constrain models and improve their realism.

The YMDF model demonstrated reasonable performance in the probed wavelength regimes while producing temporally resolved spectra with the exception for the blue continuum rise in FUV-B segment. The models calculated with the H6 and H6CII spectral setups reproduced the observed overall energy release in flares both in optical and FUV components, separately. While temporal flare models cannot fully capture the complex morphology of flares, they are still effective in reproducing the energy release across a range of flare events, particularly the Mendoza model. Our results indicate that FFDs of young M dwarfs follow a common statistical law, well-described by a two-coefficient broken power law. The YMDF module supports synthetic population generation based on these findings, with flexibility to modify the setup as needed.

Future work will apply YMDF to simulate the effects of flares, especially mid-size, on planetary atmospheres through detailed chemical kinetics modelling under high-energy irradiation. This will clarify the impacts on atmospheric composition and loss processes. Additionally, the YMDF framework will be potentially extended to model stellar activity beyond young early-type M dwarfs.

\section*{Data Availability}
The data used for this study are publicly available in the the Mikulski Archive for Space Telescopes (MAST\footnote{http://archive.stsci.edu}). The code, produced by this study, is made publicly accessible in the GitHub (\url{https://github.com/cepylka/YMDF}).

\begin{acknowledgements}
We acknowledge financial support from the Research Council of Norway (RCN), through its Centres of Excellence funding scheme, projects number 332523 (PHAB, Centre for Planetary Habitability) and number 262622 (RoCS, Rosseland Centre for Solar Physics). We thank James Davenport (University of Washington, USA) and Adalyn Gibson (University of Colorado Boulder, USA) for the fruitful discussion.We thank an anonymous referee for helpful comments and critiques, which have improved the quality of this manuscript.
\end{acknowledgements}

\bibliography{bibliography}
\newpage
\begin{appendix}
\section{Detailed Methodology}
\label{sec:appendix1}

\subsection{Radiative transfer code RH1.5D}
\label{sec:appendixRT}
RH1.5D allows to calculate synthetic disk-centre intensity $I_{\nu}$, but in order to calculate monochromatic flux according to
\begin{equation}
F_{\nu}(z)=2\pi\int_{0}^{1} I_{\nu}\mu d{\mu},
\end{equation}
 where $\mu=\cos\theta$, one should calculate $I_{\nu}(\mu)$. The previous version of the RH radiative transfer code allowed for calculating that for $\mu$ with weights, so flux can be calculated as follows:
\begin{equation}
 F_{\nu} = 2\pi \int_{-1}^{1} I_{\nu}\mu d\mu$ as  $F_{\nu} = 2\pi\frac{b-a}{2}(\omega_1 I_{\nu}(1)\mu_1+\omega_2 I_{\nu}(2)\mu_2+...)
\end{equation}
We calculate $I_{\nu}(\mu)$ in order to obtain the $\mu$-dependent intensity using the following approach. Specifically, we employed parameters that can be derived with RH1.5D, such as source function $S_{\nu}$ and total opacity $\chi$:
\begin{equation}
I_0(\mu)= \frac{1}{\mu}\int^\infty_0 (S_{\nu}(\tau'_{\nu})\exp^{-\tau'_{\nu}/\mu}d\tau'_{\nu}.
\end{equation}

\subsection{TESS data and fitting}
\label{sec:appendixTBF}
The TESS electron rate $F_{e^-/\mathrm{s}}$ converts to physical flux through a two-step photometric calibration process \citep{2015ApJ...809...77S}. The instrumental magnitude is derived from electron counts:
\begin{equation}\label{eq:91}
T_{\mathrm{mag}} = -2.5 \log_{10}(F_{e^-/\mathrm{s}}) + 20.44
\end{equation}
The TESS photometric system is anchored to the Vega-relative Cousins I-band, with its zero-point flux density defined as
\begin{equation}
F_\nu^{\mathrm{Jy}} = 2416 \times 10^{-0.4 T_{\mathrm{mag}}}
\end{equation}
The 2416 Jy value originates from the Vega flux density at the central wavelength of TESS's bandpass, $\lambda_{\mathrm{eff}}=7865,\AA$, calibrated through synthetic photometry of Vega's spectrum \citep{2015ApJ...809...77S}.
To approximate the bandpass-integrated flux $F_{\mathrm{erg,s^{-1},cm^{-2}}}$, the effective wavelength serves as a spectral anchor for unit conversion:
\begin{equation}
F_{\mathrm{erg,s^{-1},cm^{-2}}} \approx F_\nu^{\mathrm{Jy}} \times 10^{-23} \times \frac{c}{\lambda_{\mathrm{eff}}^2} \times \Delta\lambda_{\mathrm{TESS}}
\end{equation}
where $c$ is the speed of light and $\Delta\lambda_{\mathrm{TESS}} \approx 4000,\AA$ represents the approximate bandpass width. This approximation assumes wavelength-independent flux density across the bandpass. The final quantity represents the total flux integrated over TESS's 6000–10000 $\AA$ sensitivity range.

We convert all model fluxes (in cgs units) using a simple approximation for the T0 Vega following \citep{2015ApJ...809...77S}. It is convenient to define a star's TESS magnitude, assuming Vega has a flux density of $F_\lambda$=4.03$\times$10$^{-6}$, erg s$^{-1}$ cm$^{-2}$ $\AA^{-1}$ at T=0. Then, using the Eq.~\ref{eq:91}, we convert the magnitudes to electron counts per second.

For several flares in our TESS sample we additionally provide the observational data and stated that in Table~\ref{tab:102}.

\begin{table}
\scriptsize
\caption{Observational data for flares in TESS depicted in Fig. \ref{fig:87}, upper row} \label{tab:102}
\centering
\resizebox{\columnwidth}{!}{
\begin{tabularx}{8cm}{lcccc}
Flare ID & Sector & Flare start& Flare end & ED in TESS, s \TBstrut\\
\hline
 AU Mic T1 & 27 & 2037.674 & 2037.694 & 26.6 \TBstrut\\
 AU Mic T2 & 27 & 2047.213 & 2047.232 & 11.8 \TBstrut\\
 AU Mic T3 & 27 & 2053.428 & 2053.451 & 33.0\TBstrut\\
 AU Mic T4 & 27 & 2056.912 & 2056.937 & 43.6\TBstrut\\
 HIP 107345 T1 & 68& 3174.784 & 3174.902 & 437.5\TBstrut\\
 2MASS...036 T1 &69 &3188.590 & 3188.613 & 294.0\TBstrut\\
 2MASS...036 T2 &69& 3191.328 &  3191.356 & 270.6 \TBstrut\\
 2MASS...441 T1 &27& 2046.058 &2046.079& 139.7\TBstrut\\
\hline
\hline
\end{tabularx}
}
   \tablefoot{2MASS...036 and 2MASS...441 stand for 2MASS J02365171-5203036 and 2MASS J22025453-6440441, respectively.}
\end{table}

\subsection{YMDF inverse calculations of the model setup coefficients}
\label{sec:appendixYMDFED}
For producing a synthetic temporally and spectrally resolved flare dataset, we determine the coefficients of the model components, m2F12-37-2.5 and cF13-500-3, by utilising the observed EDs and the ratios derived from our model setups. This is accomplished by integrating the contributions of each component over a broad wavelength range $F_1$ and $F_2$, as well as by integrating the preflare flux $F_0$ within the model framework, and substituting them into the ED equation (see Eq.~\ref{eq:7}):

\begin{equation}\label{eq:9}
ED = \int \left( \frac{ F_\mathrm{model}(t) \cdot \left[ c x (F_1 - F_0) + x (F_2 - F_0) \right] + F_0 }{ F_0 } - 1 \right) dt,
\end{equation}
where c is the ratio specific to the model setup, and $F_\mathrm{model}(t)$ is a temporal flare model. The full derivation is presented below:

\begin{equation}\label{eq:10}
 ED = \int \left( \frac{ F_\mathrm{model}(t) \cdot \left[ c x (F_1 - F_0) + x (F_2 - F_0) \right] }{ F_0 } \right) dt
\end{equation}

\begin{equation}\label{eq:11}
ED = x \int \left( \frac{ F_\mathrm{model}(t) \cdot \left[ c (F_1 - F_0) + (F_2 - F_0) \right] }{ F_0 } \right) dt
\end{equation}

\begin{equation}\label{eq:12}
\quad A = \frac{ c (F_1 - F_0) + (F_2 - F_0) }{ F_0 }, \qquad S = \int F_\mathrm{model}(t) \, dt
\end{equation}

\begin{equation}\label{eq:13}
ED = x \cdot A \cdot S \\[2ex]
\end{equation}

\begin{equation}\label{eq:14}
x = \frac{ED}{A \cdot S}
\end{equation}

This approach enables inverse calculations of the coefficient for the second model component (x), the first model component is calculated by multiplying it by the ratio (c $\cdot$ x), and is used in the YMDF module to produce the simulated ED distribution.

\subsection{COS data handling}
\label{sec:appendixcos}
We worked with the archival data retireved by \citetalias{2025A&A...700A..53M} from the HST-COS G130M grating, program IDs: 11533 (PI Green, 2MASS J18141047-3247344, observation ID "lb3q01060"); 14784 (PI Shkolnik, 2MASS J01521830-5950168, observation ID ldab08030; 2MASS J02001277-0840516, observation ID "ldab54030"; 2MASS J03315564-4359135, observation ID "ldab10010"; 2MASS J22025453-6440441, observation ID "ldab06030"; 2MASS J02365171-5203036, observation ID "ldab02010"); 15955 (PI Richey-Yowell, HIP-107345, observation ID "le5104030"; HIP 1993, observation ID "le5108030");
16164 (PI Cauley, AU Mic, observation IDs "lebb01010", "lebb02010", "lebb03010");
16482 (PI Roman-Duval, 2MASS J11173700-7704381, observation ID "lein2d010");
17428 (PI France, AU Mic, observation IDs "lf71z1010", "lf7101010"; Karmn J07446+035, observation IDs "lf7109010", "lf71z9010").

For AU~Mic and Karmn~J07446+035, the configuration of G130M grating provided coverage from approximately $\lambda \in$ [1060,1360]~$\AA$ with a detector gap $\lambda \in$ [1210,1225]~$\AA$, masking the bright Ly$\alpha$ emission feature to avoid detector saturation. For the rest of the stars in the sample, the wavelength coverage of this configuration is $\lambda \in$ [1170,1430]~$\AA$ and includes strong emission lines of C II, C III, Si III, Si IV, and N V formed in the stellar transition region, along with several weaker lines, including some coronal iron lines. Ly$\alpha$ and O~I also fall within the bandpass; however, they are generally obscured due to contamination from geocoronal airglow.

All data were processed with the CALCOS pipeline, and time-resolved analysis was carried out using the corrtag products with a 10~s sampling interval. During the CALCOS calibration, regions affected by bright airglow, specifically around Ly$\alpha$ and O~I, are automatically flagged and masked. We applied the same masking when integrating fluxes for light-curve fitting in both the observed data and the fitted configurations (COScut1 and COScut2), ensuring consistency between the calibrated data and our analysis. The same masking procedure was also adopted when constructing light curves for the inverse models (H6 and H6CII), and in constructing the `fiducial flare' light curves.

In Figs.~\ref{fig:15}, \ref{fig:26}, and \ref{fig:25}, the model spectra are shown with masking applied for the COScut1/COScut2 setups, but without masking in the inverse models. For direct model-observation comparisons, only continuum regions were used in the corresponding spectral images. In the integrated flux analysis (Fig.~\ref{fig:85}), identical masks were applied to both the observed and model data in each panel to ensure consistency.

\begin{table*}
\scriptsize
\caption{Observational data for flares in HST-COS FUV depicted in Fig. \ref{fig:87}, bottom row} \label{tab:101}
\centering
\resizebox{0.98\textwidth}{!}{\begin{tabularx}{18cm}{lcccccccc}
Flare ID & Proposal & PI& Observation ID & Orbit & Observation date& Flare start, UT& Flare end, UT& ED in FUV, s \TBstrut\\
\hline
 AU Mic F1 & 16164 & Cauley &lebb01010& lebb01sdq & 2021-05-28 & 07:02:03 & 07:09:53 & 639.31 \TBstrut\\
 AU Mic F2 & 16164 & Cauley &lebb02010& lebb02nzq & 2022-10-09 & 17:33:15 & 17:37:05 & 335.65\TBstrut\\
 AU Mic F3 & 16164 & Cauley &lebb02010& lebb02o1q & 2022-10-09 & 19:32:00 & 19:33:40 & 156.04\TBstrut\\
 AU Mic F4 & 16164 & Cauley &lebb03010& lebb03xxq & 2021-09-24 & 21:40:06 & 21:41:46 & 72.36\TBstrut\\
 2MASS J01521830-5950168 F1 & 14784& Shkolnik &ldab08030&  ldab08wxq & 2017-08-17 & 23:38:47 &23:41:32& 26.72\TBstrut\\
 Karmn J07446+035 F1 & 17428 & France &lf71z9010&  lf71z9hiq & 2024-09-21 & 17:46:08 & 17:50:08 & 637.88\TBstrut\\
 2MASS J02365171-5203036 F1 &14784& Shkolnik &ldab02010&  ldab02mpq & 2017-08-09 & 13:24:47 &13:37:08& 591.12 \TBstrut\\
 2MASS J22025453-6440441 F1 &14784& Shkolnik &ldab06030&  ldab06n9q & 2017-08-30 & 06:58:28 &06:59:28& 8.69\TBstrut\\
\hline
\hline
\end{tabularx}
}
\end{table*}

\subsection{Electron beam model components}
\label{sec:appendixfit}

We reported the ratios between the coefficients in the H6 and H6CIII setups derived from the TESS and FUV bandpasses (as shown in Fig.~\ref{fig:4} and listed in Table~\ref{tab:2}). The ratios obtained from the HST-COS spectra (COScut1 and COScut2 setups) are systematically larger. We suggested that this issue can be resolved by including a more extensive list of spectral lines in the RH1.5 calculations. Additionally, this indicates differing behaviour among the model components. For broadband energy estimates based on model spectra, the coefficient ratio corresponding to the TESS bandpass provides an accurate description of the flare energetics across a wide wavelength range. In contrast, the HST-COS-based ratios tend to overestimate the relative contribution of the m2F12-37-2.5 component, likely due to the narrower spectral coverage and the dominance of this energetic beam model in the FUV regime. A plausible interpretation is that the HST-COS G130M grating samples the spectral region where the m2F12-37-2.5 model predicts the strongest flux, reaching values of 10$^{6}$-10$^{9}$ erg~cm$^{-2}$~s$^{-1}$ under the idealized assumption of the entire stellar surface participating in the flare (see Fig.~\ref{fig:6}). The comparison further suggests a physical distinction between the spectral domains. While in the TESS bandpass, reproducing the observed continuum rise requires the inclusion of a persistent, high-energy electron beam (represented by the cF13-500-3 component), and in the FUV, the emission is better explained by the impulsive, lower-flux beam dominating m2F12-37-2.5 model, the optical and far-ultraviolet ranges better represent different processes in flares, namely ribbons and kernels formation on the stellar surface, as we mentioned in Sect.~\ref{subsec:steatmodel}.

\subsection{Temporal flare models: further considerations}
\label{sec:appendixfit}

\begin{figure*}
\sidecaption
\includegraphics[width=18cm]{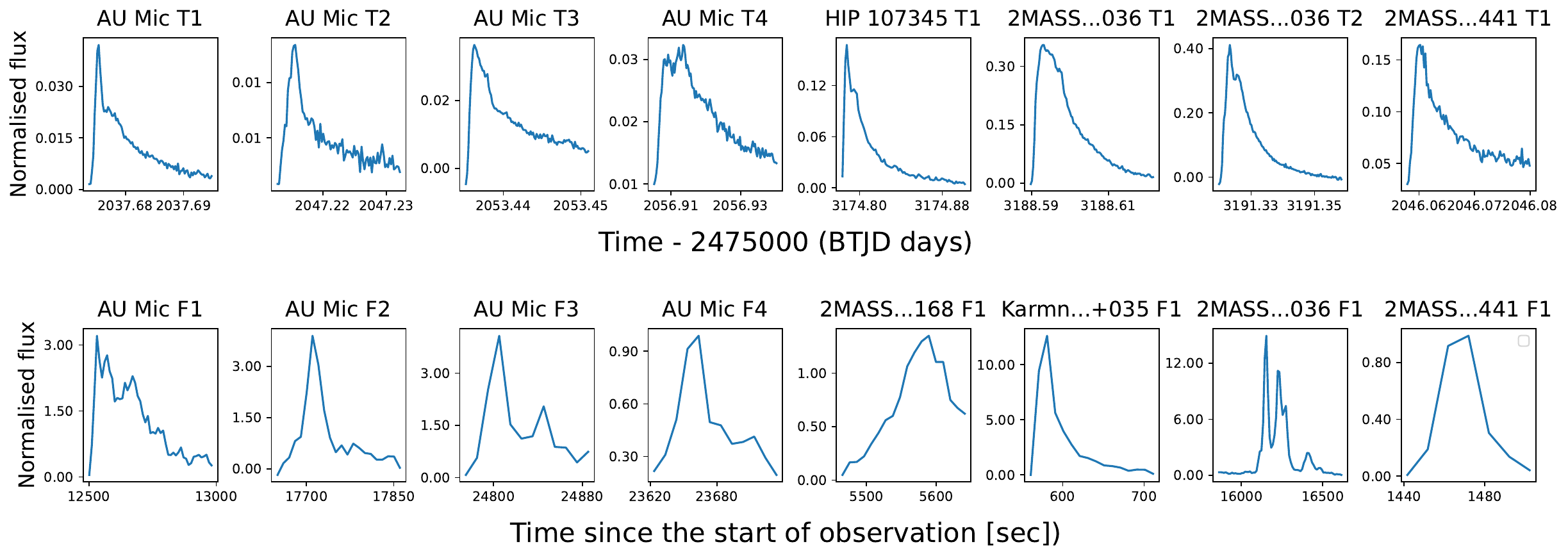}
\caption{Morphology of several stellar flares observed in TESS (upper row) and HST-COS FUV (lower row). The upper row show light curves obtained with TESS and the details of observations for these flares can be found in Table~\ref{tab:102}. In the panel labels, 2MASS...036 and 2MASS...441 stand for 2MASS~J02365171-5203036 and 2MASS~J22025453-6440441, respectively. The lower row depict light curves produced by integrating observed flux over the wavelength range in FUV with HST-COS, the details of observations for these flares can be found in Table~\ref{tab:101}. In the panel labels, 2MASS...168, Karmn...+035, 2MASS...036 and 2MASS...441 stand for 2MASS~J01521830-5950168, Karmn~J07446+035, 2MASS~J02365171-5203036 and 2MASS~J22025453-6440441, respectively.}

\label{fig:87}
\end{figure*}

\begin{table*}
\scriptsize
\caption{The time-integrated flare flux in electron counts (e$^-$) in TESS} \label{tab:10}
\centering
\resizebox{0.99\textwidth}{!}{\begin{tabularx}{18cm}{lllllllll}
\hline
Temporal flux evolution & AU Mic T1 & AU Mic T2 & AU Mic T3 & AU Mic T4 & HIP 107345 T1 & 2MASS...036 T1 & 2MASS...036 T2 & 2MASS...441 T1\TBstrut\\

\hline
Mendoza model & 4.9236$\times$10$^{8}$ & 4.3249$\times$10$^{8}$ & 5.5544$\times$10$^{8}$ & 6.2689$\times$10$^{8}$ & 1.4789$\times$10$^{8}$ & 3.6928$\times$10$^{7}$ & 4.3043$\times$10$^{7}$ & 1.7997$\times$10$^{7}$ \TBstrut\\
Davenport model & 4.9240$\times$10$^{8}$ & 4.3249$\times$10$^{8}$ & 5.5546$\times$10$^{8}$ & 6.2700$\times$10$^{8}$ & 1.4796$\times$10$^{8}$ & 3.6952$\times$10$^{7}$ & 4.3091$\times$10$^{7}$ & 1.8003$\times$10$^{7}$ \TBstrut\\
Feinstein model & 4.8899$\times$10$^{8}$ & 4.3130$\times$10$^{8}$ & 5.5058$\times$10$^{8}$ & 6.2088$\times$10$^{8}$ & 1.4306$\times$10$^{8}$ & 3.5314$\times$10$^{7}$ & 4.1490$\times$10$^{7}$ & 1.7444$\times$10$^{7}$ \TBstrut\\
Inverse H model & 4.9266$\times$10$^{8}$ & 4.3275$\times$10$^{8}$ & 5.5578$\times$10$^{8}$ & 6.2726$\times$10$^{8}$ & 1.4858$\times$10$^{8}$ & 3.6747$\times$10$^{7}$ & 4.2832$\times$10$^{7}$ & 1.7973$\times$10$^{7}$ \TBstrut\\
Inverse HCIII model & 4.9266$\times$10$^{8}$ & 4.3275$\times$10$^{8}$ & 5.5577$\times$10$^{8}$ & 6.2726$\times$10$^{8}$ & 1.4858$\times$10$^{8}$ & 3.6747$\times$10$^{7}$ & 4.2832$\times$10$^{7}$ & 1.7972$\times$10$^{7}$ \TBstrut\\
`Fiducial flare' model & 4.5203$\times$10$^{8}$ & 3.6967$\times$10$^{8}$ & 4.8119$\times$10$^{8}$ & 5.3688$\times$10$^{8}$ & - & - & - & - \TBstrut\\
Observed & 4.9235$\times$10$^{8}$ & 4.3267$\times$10$^{8}$ & 5.5564$\times$10$^{8}$ & 6.2805$\times$10$^{8}$ & 1.4729$\times$10$^{8}$ & 3.6497$\times$10$^{7}$ & 4.2349$\times$10$^{7}$ & 1.8054$\times$10$^{7}$ \TBstrut\\
\hline
\end{tabularx}
}
   \tablefoot{The time-integrated flare flux is calculated for [6000,1000] $\AA$ range in all temporally-spectrally-resolved models, including `fiducial flare'. 2MASS...036 and 2MASS...441 stand for 2MASS J02365171-5203036 and 2MASS J22025453-6440441, respectively.}
\end{table*}

\begin{table*}
\scriptsize
\caption{The time-integrated flare flux (erg cm$^{-2}$) in HST-COS FUV} \label{tab:11}
\centering
\resizebox{0.99\textwidth}{!}{\begin{tabularx}{18cm}{lllllllll}
\hline
Temporal flux evolution& AU Mic F1 & AU Mic F2 & AU Mic F3 & AU Mic F4 & 2MASS...168 F1 & Karmn...+035 F1 & 2MASS...036 F1 & 2MASS...441 F1 \TBstrut\\
\hline
Mendoza model & 1.2787$\times$10$^{-9}$ & 7.0657$\times$10$^{-10}$ & 3.2834$\times$10$^{-10}$ & 1.8946$\times$10$^{-10}$ & 8.8510$\times$10$^{-11}$ & 7.1278$\times$10$^{-10}$ & 1.5559$\times$10$^{-9}$ & 6.3635$\times$10$^{-11}$ \TBstrut\\
Davenport model & 1.2769$\times$10$^{-9}$ & 7.1558$\times$10$^{-10}$ & 3.2823$\times$10$^{-10}$ & 1.9141$\times$10$^{-10}$ & 8.9296$\times$10$^{-11}$ & 7.2030$\times$10$^{-10}$ & 1.6056$\times$10$^{-9}$ & 6.5371$\times$10$^{-11}$ \TBstrut\\
Feinstein model & 6.3695$\times$10$^{-10}$ & 4.2002$\times$10$^{-10}$ & 2.2379$\times$10$^{-10}$ & 2.1785$\times$10$^{-10}$ & 8.1210$\times$10$^{-11}$ & 4.0197$\times$10$^{-10}$ & 7.3248$\times$10$^{-10}$ & 6.0898$\times$10$^{-11}$ \TBstrut\\
Inverse H model & 1.2908$\times$10$^{-9}$ & 6.4422$\times$10$^{-10}$ & 3.1229$\times$10$^{-10}$ & 2.2236$\times$10$^{-10}$ & 8.6776$\times$10$^{-11}$ & 5.4157$\times$10$^{-10}$ & 1.6026$\times$10$^{-9}$ & 7.2716$\times$10$^{-11}$ \TBstrut\\
Inverse HCIII model & 1.4917$\times$10$^{-9}$ & 7.1842$\times$10$^{-10}$ & 3.5056$\times$10$^{-10}$ & 2.7133$\times$10$^{-10}$ & 9.2140$\times$10$^{-11}$ & 5.3430$\times$10$^{-10}$ & 1.4785$\times$10$^{-9}$ & 7.2198$\times$10$^{-11}$ \TBstrut\\
`Fiducial flare' model & 2.5810$\times$10$^{-9}$ & 8.4980$\times$10$^{-10}$ & 5.0288$\times$10$^{-10}$ & 3.5072$\times$10$^{-10}$ & - & - & - & - \TBstrut\\
Observed & 1.2793$\times$10$^{-9}$ & 6.7371$\times$10$^{-10}$ & 3.2174$\times$10$^{-10}$ & 1.9310$\times$10$^{-10}$ & 8.5826$\times$10$^{-11}$ & 5.9483$\times$10$^{-10}$ & 1.4291$\times$10$^{-9}$ & 6.8553$\times$10$^{-11}$ \TBstrut\\

\hline
\end{tabularx}
}
   \tablefoot{The time-integrated flare flux is calculated across FUV-A and FUV-B segments consistent with detector gaps specific to the observation (see Appendix~\ref{sec:appendixcos}) in all temporally-spectrally-resolved models, including `fiducial flare'. 2MASS...168, Karmn...+035, 2MASS...036 and 2MASS...441 stand for 2MASS J01521830-5950168, Karmn J07446+0352MASS, J02365171-5203036 and 2MASS J22025453-6440441, respectively.}
\end{table*}

As previously discussed in Section~\ref{subsec:sydi}, the Mendoza model provides a better match to the total time-integrated flux of the flare light curve, accurately reproducing the overall energy output. In practical terms, this means that when integrating the flare emission over time, the total energetic budget is very well recovered. However, it typically underestimates the flare amplitude (see Fig.~\ref{fig:85}), which indicates that the model has difficulty reproducing the sharp, high-intensity peak of some observed events. Therefore, as we adopted the simpler Gaussian rise–exponential decay temporal flare model proposed by \citet{2022AJ....164..110F}, we suggest that with a modification allowing the rise phase to be described by a Student’s t-distribution having a non-integer fractional degrees of freedom of 2.05, this model would successfully reproduce the flare amplitude. This modification accounts for sharper peaking and better captures the observed flare rise morphology, particularly for flares with fast impulsive phases. Additionally, we set the decay parameter to $\gamma=-0.09$, which corresponds to a slower decay in time and produces a better match to the gradual phase in some flares.

However, the complex morphology of observed flares suggests that this component should be modified to reflect the evolution of individual flares. This modification is not yet implemented in the YMDF module, partly because our approach cannot currently be grounded in statistical studies of flare temporal profiles, which are absent and remain a subject for future work. Of all the approaches evaluated, this parametrisation yielded the best agreement with the observed flare profiles.
Although the fitted coefficients are expected to be larger when using the Feinstein model compared to the Mendoza and Davenport models, the resulting ratios of the obtained coefficients remain very similar between the two models (for 2MASS~J02365171$-$5203036 (COScut2), $\hat{X}_1/\hat{X}_2 \sim$~11.30 and for AU~Mic (COScut1), the corresponding ratios~$\sim$ 13.54, respectively).

For several stars, namely 2MASS~J02365171-5203036, 2MASS~J18141047-3247344, and Karmn~J07446+035, the Feinstein model enabled us to successfully fit the YMDF H6 model to the peak of the flare observed in the FUV HST-COS spectra (see, e.g., Fig.~\ref{fig:85} for AU Mic F1 and Karmn J07446+035 F1 panels in the bottom row), while for TESS fits it can overestimate the peak (see the same figure, AU Mic T2 and HIP 107345 T1 panels in the upper row). Although the modified Feinstein model yields a superior fit to the flare peak and enables more accurate estimation of the energy at the most energetic phase, including for relatively small flares observed in the FUV with HST-COS comparing to the TESS observations, it does not capture the full energy release within a specific wavelength range, for which the Mendoza model performs significantly better (see Tables \ref{tab:10} and \ref{tab:11}).

\label{sec:appendix2}

\begin{figure*}
\resizebox{\hsize}{!}{
   \centering
   \begin{subfigure}{}
      \includegraphics{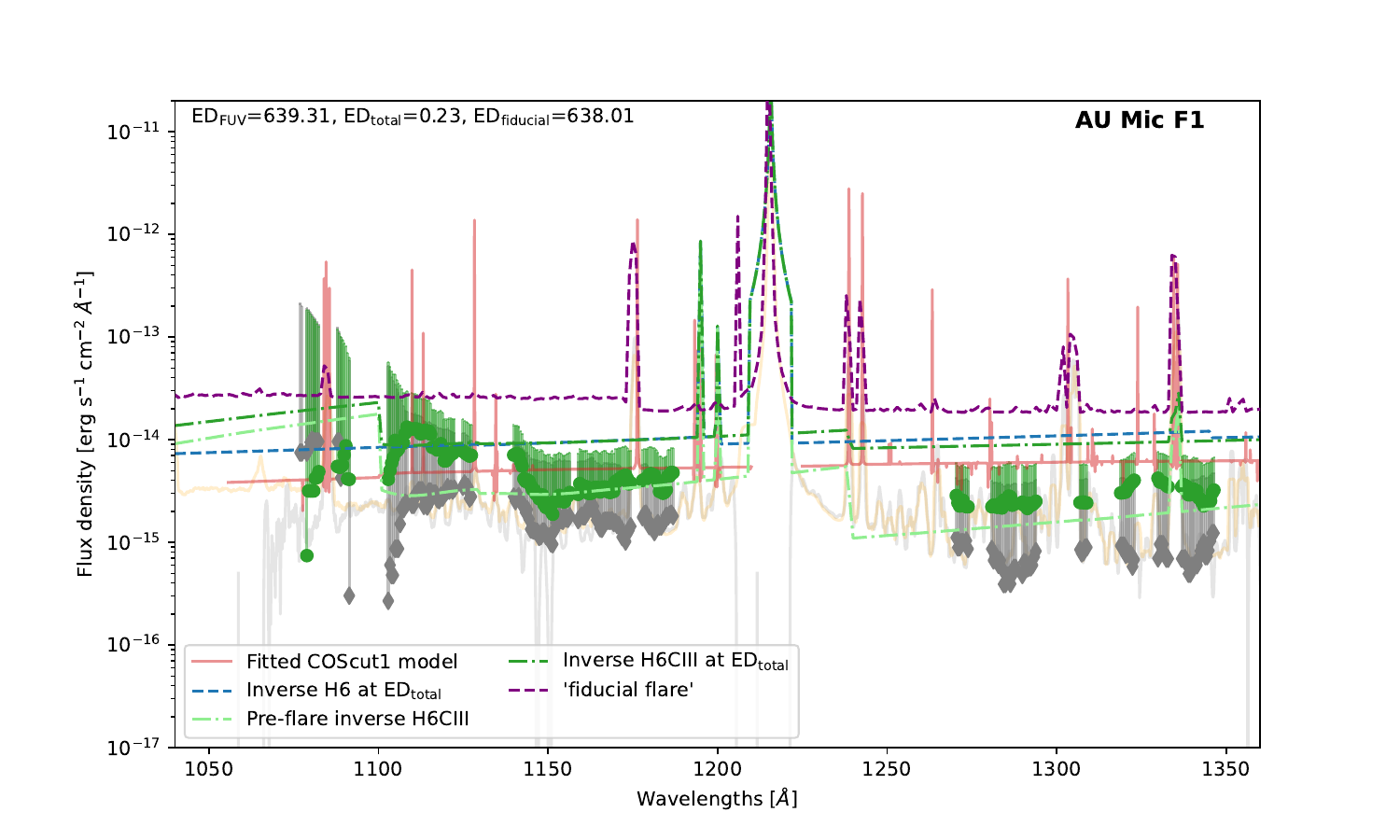}
   \end{subfigure}
   \begin{subfigure}
   \centering{}
      \includegraphics{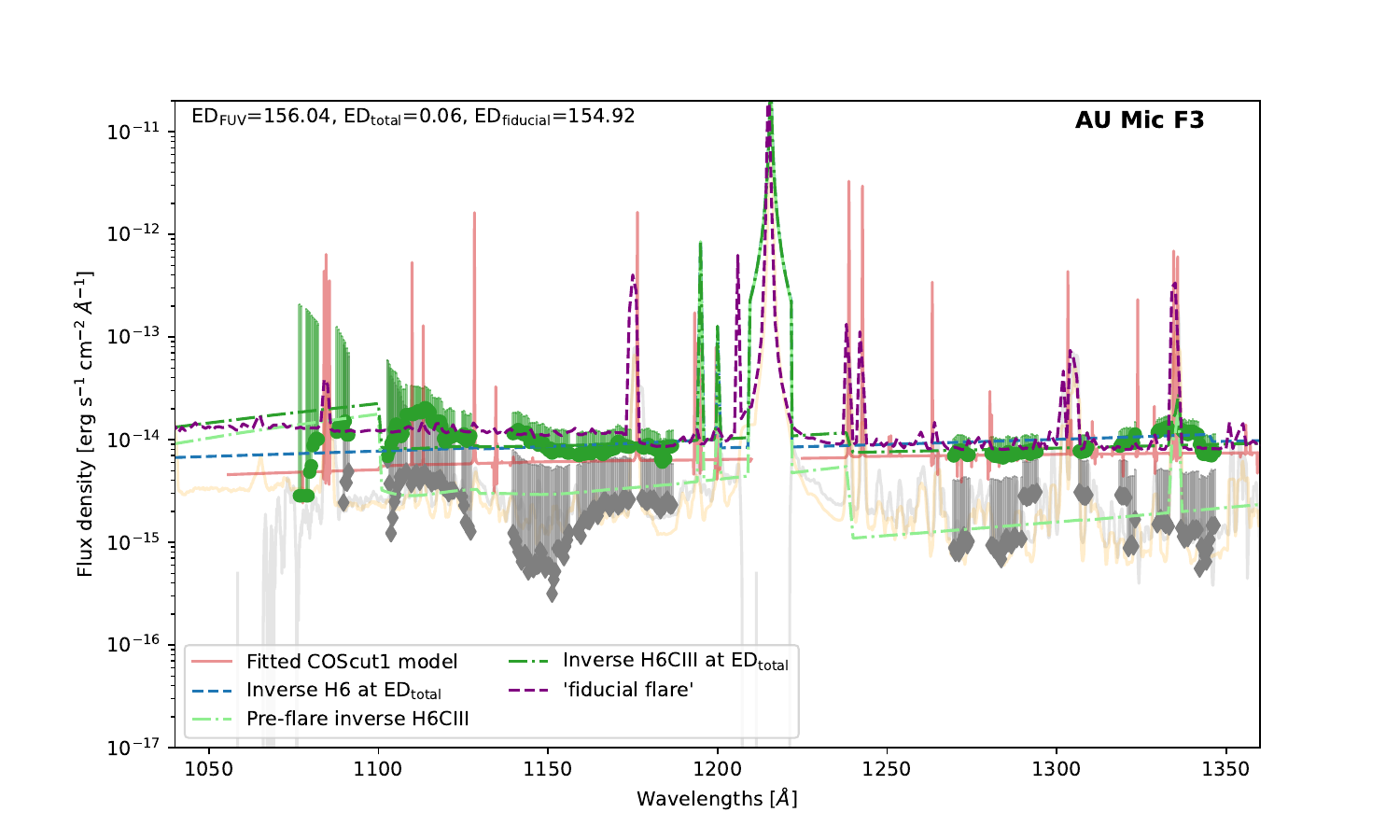}
   \end{subfigure}
   }
\caption{AU Mic spectra at the peak of flares observed with HST-COS. In both panels, green spheres denote continuum flux measurements at the flare peaks, while grey diamonds indicate the quiescent continuum levels. Associated uncertainties are depicted as light-coloured lines, green for the flare peak and grey for quiescence, with only upper error bars shown for clarity, given the symmetry of the errors. The light red lines represent the best-fit obtained using the COScut1 setup. Green and light green dashed lines correspond to the H6CIII models with $\hat{X}_1$/$\hat{X}_2$=8.74 derived from TESS photometry for the flare peak and pre-flare states, respectively, while the blue dash-dotted lines show the simplified H6 models with $\hat{X}_1$/$\hat{X}_2$=5.03. The background light grey spectra are the xdsum products for each respective orbit, representing the summed, extracted one-dimensional spectra from all exposures within an orbit and providing high signal-to-noise reference spectra for comparison. Light orange lines display smoothed panchromatic AU Mic spectra from \citet{2022AJ....164..110F}, and purple dashed lines depict the `fiducial flare' model from \citet{2018ApJ...867...71L} at similar equivalent durations, included for context.}
\label{fig:26}
\end{figure*}
\begin{figure*}\resizebox{\hsize}{!}{
   \centering
   \begin{subfigure}{}
      \includegraphics{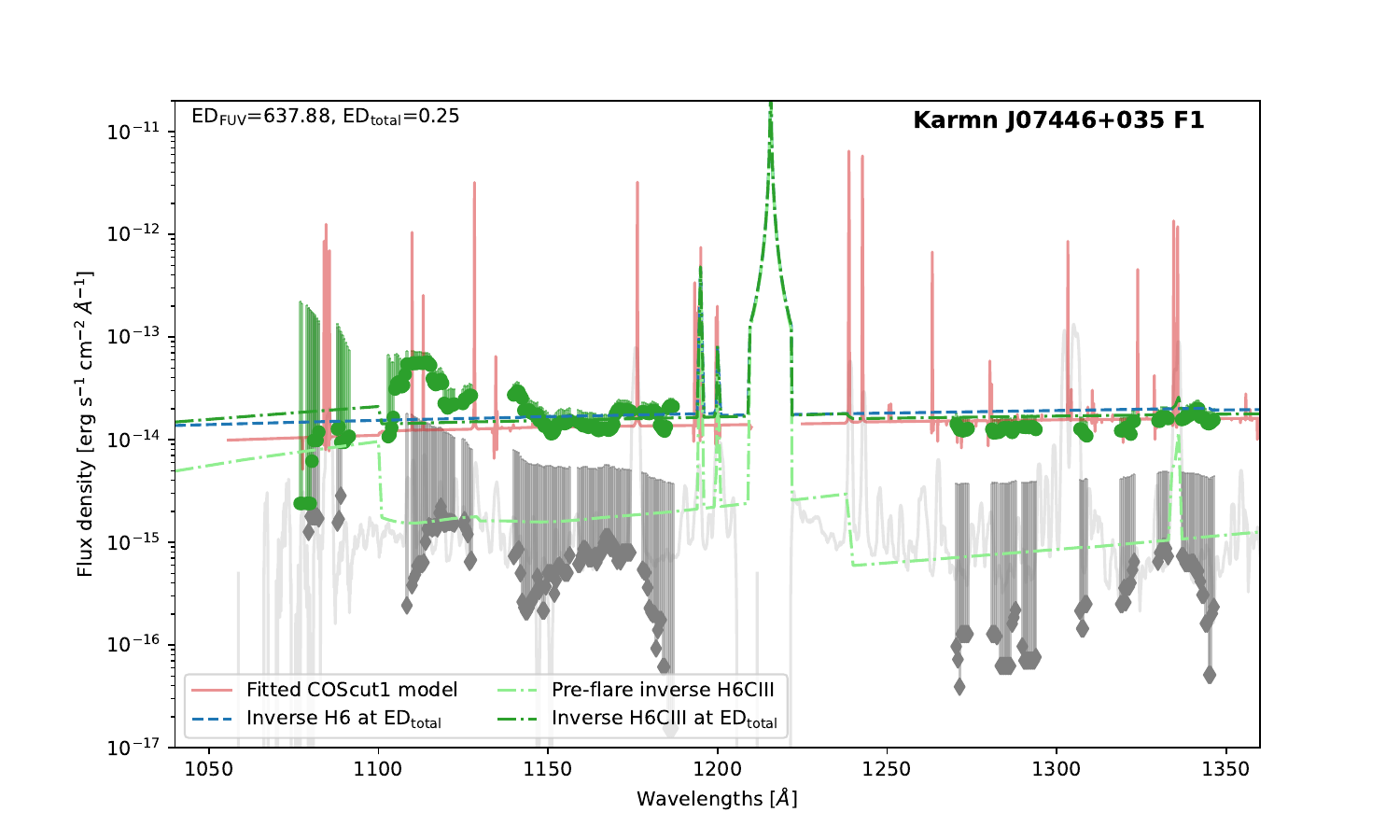}
   \end{subfigure}
   \begin{subfigure}
   \centering{}
      \includegraphics{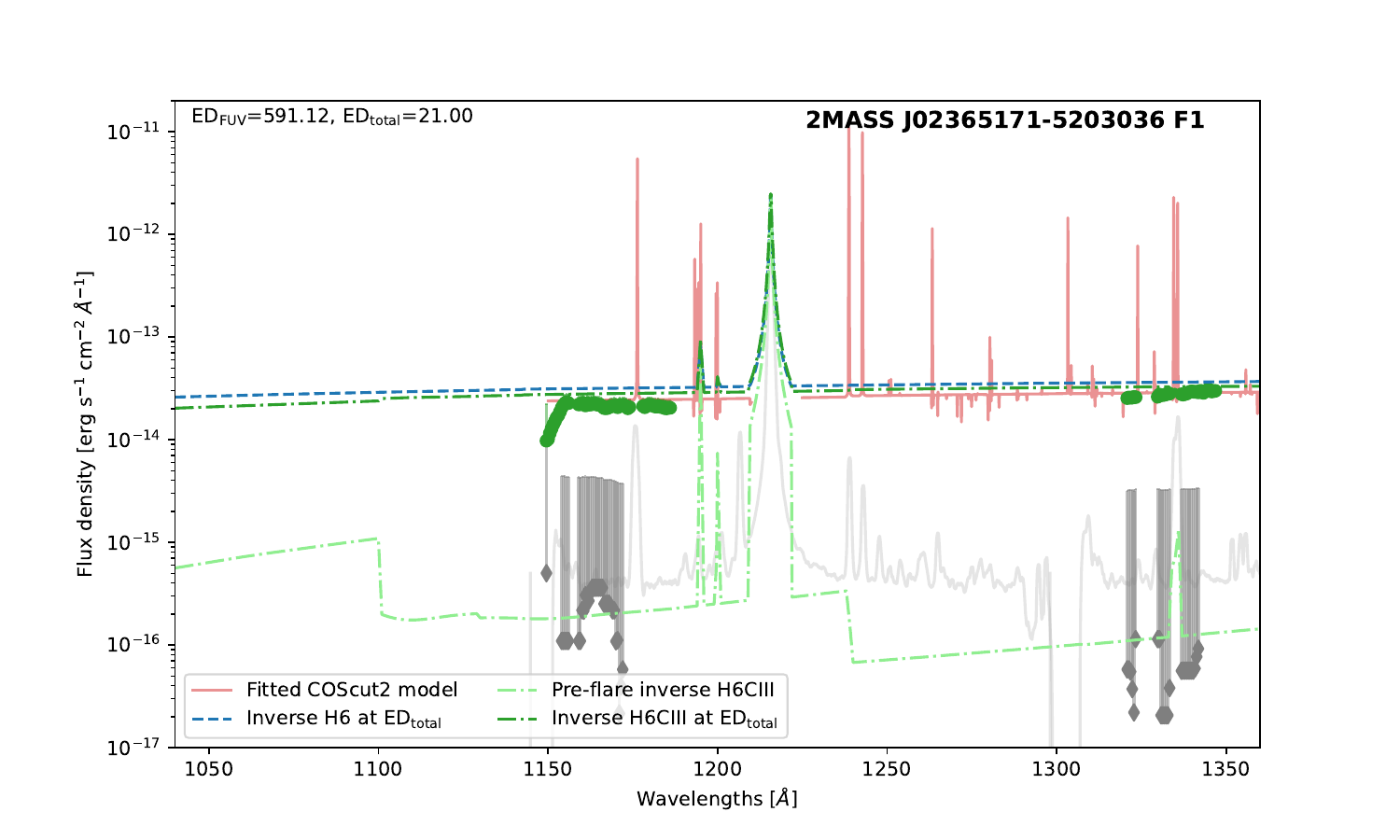}
   \end{subfigure}
   }
\caption{The spectra at the peak of flares observed with HST-COS for Karmn~J07446+035 (left) and 2MASS~J02365171-5203036 (right). For both stars, green spheres indicate continuum flux measurements at the flare peak, while grey diamonds represent the quiescent continuum levels. Associated uncertainties are shown as light-coloured lines-green for the flare peak and grey for quiescence-with only upper error bars displayed for clarity, as the errors are symmetric. The light red lines correspond to the best-fit continuum models obtained using the COScut1 (left panel) and COScut2 (right panel) setups. Green and light green dashed lines represent the inverse H6CIII models for the flare peak and pre-flare states, respectively, while blue dash-dotted lines show the simplified inverse H6 models. The background light grey spectra are the xdsum spectrum.
}
\label{fig:25}
\end{figure*}

\end{appendix}
\end{document}